\documentclass[twocolumn,numberedappendix]{emulateapj} \shorttitle{New Limits on the $21\,\textrm{cm}$ Power Spectrum at $z=8.4$}
\shortauthors{Ali, et al.}

\usepackage{amsmath} \usepackage{graphicx} \usepackage[figuresright]{rotating}
\usepackage{natbib}
 \def\k{\mathbf{k}} \def\r{\mathbf{r}} 
 
  \def\expval#1{\langle #1
\rangle}   
 
\newcommand{\x}{\mathbf{x}} 
 
\newcommand{\C}{\mathbf{C}} \newcommand{\Q}{\mathbf{Q}}

\newcommand{\phat}{\hat{\mathbf{p}}}
\newcommand{\qhat}{\hat{\mathbf{q}}}
\newcommand{\hMpci}{h\ {\rm Mpc}^{-1}}
\newcommand{\Tsys}{T_{\rm sys}}
\newcommand{\Tspin}{T_{\rm s}}
\newcommand{\kmin}{k_{\rm min}}
\newcommand{\kmax}{k_{\rm max}}
\newcommand{\Tcmb}{T_\gamma}

\newcommand{\mKlimit}{(22.4\,\textrm{mK})$^2$ }

\newcommand{\kolopaniscitet}{\textrm{Kolopanis et al. (2018, in prep)}}
\newcommand{\chengcitet}{\textrm{Cheng et al. (2018, submitted)}}
\newcommand{\chengcitep}{\textrm{(Cheng et al. 2018, submitted)}}

\newcount\colveccount
\newcommand*\colvec[1]{
        \global\colveccount#1
        \begin{pmatrix}
        \colvecnext
}
\def\colvecnext#1{
        #1
        \global\advance\colveccount-1
        \ifnum\colveccount>0
                \\
                \expandafter\colvecnext
        \else
                \end{pmatrix}
        \fi
}

\begin{document}

\title{PAPER-64 Constraints on Reionization: the  $21\,\textrm{cm}$ Power Spectrum at $z=8.4$}

\author{
Zaki S. Ali\altaffilmark{1}, 
Aaron R. Parsons\altaffilmark{1,2}, 
Haoxuan Zheng\altaffilmark{3},
Jonathan C. Pober\altaffilmark{4}, 
Adrian Liu\altaffilmark{1,5}, 
James E. Aguirre\altaffilmark{6},
Richard F. Bradley\altaffilmark{7,8,9},
Gianni Bernardi\altaffilmark{10,11,12}, 
Chris L. Carilli\altaffilmark{13,14},
Carina Cheng\altaffilmark{1},
David R. DeBoer\altaffilmark{2}, 
Matthew R. Dexter\altaffilmark{2},
Jasper Grobbelaar\altaffilmark{10},
Jasper Horrell\altaffilmark{10},
Daniel C. Jacobs\altaffilmark{15}, 
Pat Klima\altaffilmark{8},
David H. E. MacMahon\altaffilmark{2},
Matthys Maree\altaffilmark{10},
David F. Moore\altaffilmark{6},
Nima Razavi\altaffilmark{14},
Irina I. Stefan\altaffilmark{14},
William P. Walbrugh\altaffilmark{10},
Andre Walker\altaffilmark{10}
}

\altaffiltext{1}{Astronomy Dept., U. California, Berkeley CA}
\altaffiltext{2}{Radio Astronomy Lab., U. California, Berkeley CA}
\altaffiltext{3}{Dept. of Physics, Massachusetts Inst. of Tech., Cambridge MA}
\altaffiltext{4}{Physics Dept.  U. Washington, Seattle WA}
\altaffiltext{5}{Berkeley Center for Cosmological Physics, Berkeley, CA} 
\altaffiltext{6}{Dept. of Physics and Astronomy, U. Penn., Philadelphia PA} 
\altaffiltext{7}{Dept. of Electrical and Computer Engineering, U. Virginia, Charlottesville VA}
\altaffiltext{8}{National Radio Astronomy Obs., Charlottesville VA}
\altaffiltext{9}{Dept. of Astronomy, U. Virginia, Charlottesville VA}
\altaffiltext{10}{Square Kilometer Array, S. Africa, Cape Town South Africa}
\altaffiltext{11}{Dept. of Physics and Electronics, Rhodes University}
\altaffiltext{12}{Harvard-Smithsonian Cen. for Astrophysics, Cambridge MA}
\altaffiltext{13}{National Radio Astronomy Obs., Socorro NM}
\altaffiltext{14}{Cavendish Lab., Cambridge UK}
\altaffiltext{15}{School of Earth and Space Exploration, Arizona State U., Tempe AZ}

\begin{abstract}
In this paper, we report new limits on 21cm emission from cosmic reionization
based on a 135 day observing campaign with a 64-element deployment of the
Donald C. Backer Precision Array for Probing the Epoch of Reionization (PAPER)
in South Africa.  This work extends the work presented in
\citet{parsons_et_al2014} with more collecting area, a longer observing period, improved redundancy-based
calibration, improved fringe-rate filtering, and updated power-spectral
analysis using optimal quadratic estimators. The result is a new $2\sigma$
upper limit on $\Delta^2(k)$ of \mKlimit in the range
$0.15<k<0.5\hMpci$ at $z=8.4$.  This represents a three-fold improvement over the
previous best upper limit.  As we discuss in more depth in a forthcoming paper
\citep{pober_et_al2015}, this upper limit supports and extends previous
evidence against extremely cold reionization scenarios.  We conclude with a
discussion of implications for future 21cm reionization experiments, including
the newly funded Hydrogen Epoch of Reionization Array. \\ 
\textbf{The limits presented in this paper have been retracted: The erratum can be found in Appendix \ref{app:erratum}.}
\end{abstract}

\section{Introduction}

The {\it cosmic dawn} of the universe, which begins with the birth of the first stars and ends approximately
one billion years later with the full
reionization of the intergalactic medium (IGM), represents one of the last 
unexplored phases in cosmic history. 
Studying the formation of the first galaxies and their influence on the primordial IGM during this
period is among the highest priorities in modern astronomy.
During our cosmic dawn, IGM characteristics depend on the matter density field, the mass and clustering of 
the first galaxies \citep{lidz_et_al2008}, their ultraviolet luminosities \citep{mcquinn_et_al2007},
the abundance of X-ray sources and other sources of heating \citep{pritchard_loeb2008,mesinger_et_al2013},
and higher-order cosmological effects like the relative velocities of baryons
and dark matter \citep{mcquinn_oleary2012,visbal_et_al2012}.

Recent measurements
have pinned down the bright end of the galaxy luminosity function
at $z \la 8$ \citep{bouwens_et_al2010,schenker_et_al2013} and have detected a few sources at even greater
distances \citep{ellis_et_al2013,oesch_et_al2013}. 
In parallel, a number of indirect techniques have constrained the evolution of the neutral fraction
with redshift. Examples include integral constraints on reionization from the
optical depth of Thomson scattering to the CMB \citep{planck_et_al2014,planck_et_al2015},
large-scale CMB polarization anisotropies \citep{page_et_al2007}, and
secondary temperature fluctuations generated by the kinetic Sunyaev-Zel'dovich effect \citep{mesinger_et_al2012,zahn_et_al2012,battaglia_et_al2013,park_et_al2013,george_et_al2014}.
Other probes of the tail end of reionization include
observations of resonant scattering of Ly$\alpha$ by the neutral IGM toward
distant quasars (the `Gunn-Peterson' effect) \citep{fan_et_al2006},
the demographics of Ly$\alpha$ emitting galaxies \citep{schenker_et_al2013,treu_et_al2013,Faisst_et_al2014},
and the
Ly$\alpha$ absorption profile toward very distant quasars \citep{bolton_et_al2011,bosman_becker2015}.
As it stands, the known population of galaxies falls well short 
of the requirements for reionizing the universe at redshifts compatible
with CMB optical depth measurements \citep{robertson_et_al2013,robertson_et_al2015}, 
driving us to deeper observations with, e.g., JWST and ALMA, to reveal the fainter end of the luminosity function.

Complementing these probes of our cosmic dawn are experiments targeting
the $21\,\textrm{cm}$ ``spin-flip" transition of neutral hydrogen at high redshifts.
This signal has been recognized as a potentially powerful probe
of the cosmic dawn \citep{furlanetto_et_al2006,morales_wyithe2010,pritchard_loeb2012} that can reveal
large-scale fluctuations in the ionization state and temperature of the IGM, opening
a unique window into the complex astrophysical interplay between the first luminous
structures and their surroundings.
Cosmological redshifting maps 
each observed frequency with a particular emission time (or distance), enabling $21\,\textrm{cm}$ experiments 
to eventually reconstruct 
three-dimensional pictures of the time-evolution of large scale structure in the universe. 
While such maps can potentially probe nearly the entire observable universe \citep{mao_et_al2008},
in the near term, $21\,\textrm{cm}$ cosmology experiments are focusing on statistical measures
of the signal.


There are two complementary experimental approaches to accessing $21\,\textrm{cm}$ emission from
our cosmic dawn.  So-called ``global" experiments such as 
EDGES \citep{bowman_et_al2010}, 
the LWA \citep{ellingson_et_al2013},
LEDA \citep{greenhill_bernardi2012,bernardi_et_al2015}, 
DARE \citep{burns_et_al2012}, 
SciHi \citep{tabitha_et_al2014}, 
BigHorns \citep{sokolowski_et_al2015},
and SARAS \citep{patra_et_al2015} 
seek to measure the
mean brightness temperature of $21\,\textrm{cm}$ relative to the CMB background. These experiments
typically rely on auto-correlations from a small number of dipole elements to access
the sky-averaged $21\,\textrm{cm}$ signal, although recent work is showing
that interferometric cross-correlations may also be used to access the signal
\citep{vedantham_et_al2015,presley_et_al2015}.
In contrast, experiments targeting statistical power-spectral measurements of the $21\,\textrm{cm}$
signal employ larger interferometers.  Examples of such interferometers targeting
the reionization signal include
the GMRT \citep{paciga_et_al2013},
LOFAR \citep{van_haarlem_et_al2013},
the MWA \citep{tingay_et_al2013},
the 21CMA \citep{peterson_et_al2004,wu2009},
and the Donald C. Backer Precision Array for Probe the Epoch of Reionization (PAPER; \citealt{parsons_et_al2010}). 

PAPER is unique for being a dedicated instrument with the flexibility
to explore non-traditional experimental approaches, and is converging on a self-consistent
approach to achieving both the level of foreground removal and the sensitivity that are required 
to detect the 21cm reionization signal.  This approach focuses on spectral smoothness as the primary
discriminant between foreground emission and the 21cm reionization signal  and applies an understanding
of interferometric responses in the delay domain to identify bounds on instrumental chromaticity 
(\citealt{parsons_et_al2012b}, hereafter P12b).  This type of ``delay-spectrum'' analysis permits data from each 
interferometric baseline
to be analyzed separately without requiring synthesis imaging for foreground removal.  As a result, PAPER has
been able to adopt new antenna configurations that are densely packed and highly redundant.
These configurations are poorly suited for synthesis imaging but
deliver a substantial sensitivity boost for power-spectral measurements that are not yet limited by
cosmic variance (\citealt{parsons_et_al2012a}, hereafter P12a).  Moreover, they are particularly suited
for redundancy-based calibration \citep{wieringa1992,liu_et_al2010,zheng_et_al2014}, on which PAPER
now relies to solve for the majority of the internal instrumental degrees of freedom (dof).  The efficacy of
this approach was demonstrated with 
data from a 32-antenna deployment of PAPER, which achieved an upper 
limit on the $21\,\textrm{cm}$ power spectrum $\Delta^2(k)\leq (41\,\textrm{mK})^{2}$ at 
$k=0.27\hMpci$ (\citealt{parsons_et_al2014}, hereafter P14).  That upper limit improved
over previous limits by orders of magnitude, showing that the early universe was heated
from adiabatic cooling, presumably by emission from high-mass X-ray binaries or mini-quasars.

In this paper, we improve on this previous result using
a larger 64-element deployment of PAPER and a longer observing period, along with better redundant calibration, an improved fringe-rate filtering technique, and an updated power-spectrum estimation pipeline.
The result is an
upper limit on $\Delta^2(k)$ of \mKlimit~in the range
$0.15<k<0.5\hMpci$ at $z=8.4$.  This result places constraints on the 
spin temperature of the IGM, and as is shown in a forthcoming paper,
\citet{pober_et_al2015}, this supports and extends
previous evidence against extremely cold reionization scenarios.
In Section
\ref{sec:observations} we describe the observations used in this analysis. In
Sections \ref{sec:calib} and \ref{sec:instrument}, 
we discuss the calibration and 
the stability of the PAPER instrument.
We then move on to a discussion of our power-spectrum analysis pipeline in Section
\ref{sec:oqe}. 
We present our results in Section
\ref{sec:results} along with new constraints on the 21cm power spectrum.
We discuss these results in Section \ref{sec:discussion} and conclude in Section \ref{sec:conclusion}.

\section{Observations}\label{sec:observations}

\begin{figure}\centering
\includegraphics[width=\columnwidth]{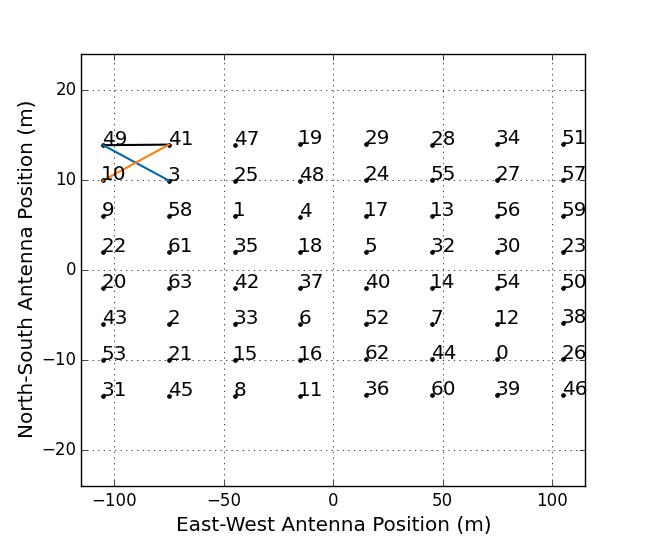}
\caption{
Antenna position within the PAPER-64 array.
This analysis only makes use of
east-west baselines between adjacent columns that have row
separations of zero (black; e.g. 49-41, 41-47, 10-3, \ldots)
one in the northward direction (orange; e.g. 10-41, 3-47, 9-3, \ldots) or
one in the southward direction (blue; e.g. 49-3, 41-25, 10-58, \ldots).
Because of their high levels of redundancy, 
these baselines constitute the bulk of the array's sensitivity for power
spectrum analysis.}
\label{fig:antenna_positions}
\end{figure}

\begin{figure*}\centering
\includegraphics[width=2\columnwidth]{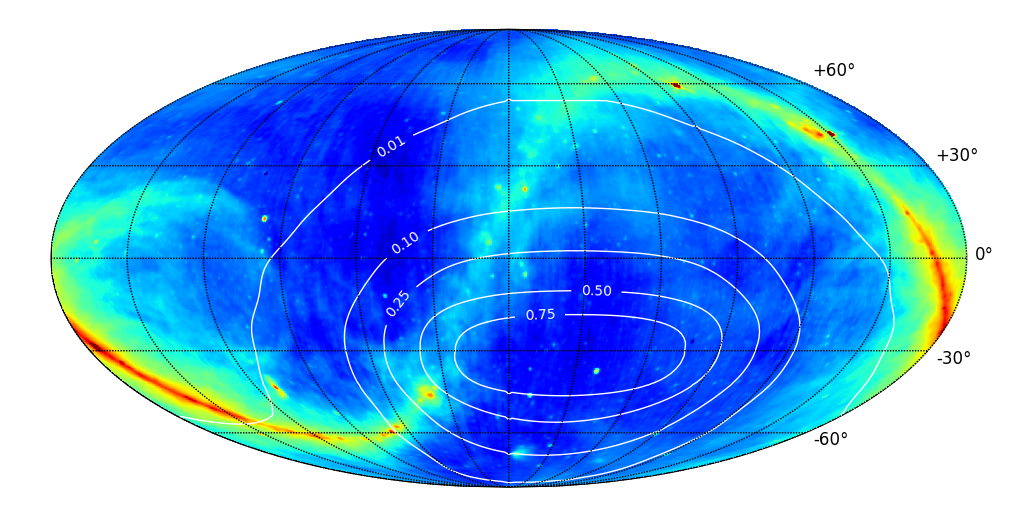}
\caption{The Global Sky Model \citep{deoliveira2008}, illustrating foregrounds to the 21cm
cosmological signal, with 
contours indicating beam-weighted observing time (relative to peak) for the PAPER observations
described in Section \ref{sec:observations}.  The map is centered at 6:00 hours in RA.
}\label{fig:coverage}
\end{figure*}

We base our analysis on drift-scan observations 
with 64 dual-polarization PAPER antennas (hereafter, ``PAPER-64") deployed 
at the Square Kilometre Array South Africa
(SKA-SA) reserve in the Karoo desert in South Africa
(30:43:17.5$^\circ$ S, 21:25:41.8$^\circ$ E).
Each PAPER element features a crossed-dipole design measuring two
linear (X,Y) polarizations.
The design of the PAPER element, 
which features spectrally and spatially smooth responses 
down to the horizon with a FWHM of $60^{\circ}$, is summarized in \citet{parsons_et_al2010}
and \citet{pober_et_al2012}.  
For this analysis, we use only the XX and YY polarization cross-products.

As shown in Figure \ref{fig:antenna_positions}, PAPER-64 employs
a highly redundant antenna layout where multiple baselines measure
the same Fourier mode on the sky (P12a; P14).
We rely on all 2016 baselines for calibration,
but only use a subset of the baselines for the power spectrum
analysis. This subset consists of three types of baselines: the 30-m
strictly east-west baselines between adjacent columns (e.g. 49-41, black 
in Figure \ref{fig:antenna_positions}; hereafter referred to 
as {\it fiducial baselines}), 30-m east-west baselines
whose eastern element is staggered one row up (e.g. 10-41, orange in Figure \ref{fig:antenna_positions}), and
those whose eastern element is one row down (e.g. 49-3, blue in Figure \ref{fig:antenna_positions}).
These baseline groups consist of 56, 49, and 49 baselines, respectively.
We define a redundant group of
baselines as being the set of baselines that have the same grid spacing;
baselines in each
of the three redundant groups described above are instantaneously redundant and
therefore measure the same Fourier modes on the sky. Thus, within a redundant group,
measurements from baselines may be 
coherently added to build power-spectrum sensitivity as $N$ rather than
$\sqrt{N}$, where $N$ is the number of baselines added.  

PAPER-64 conducted nighttime observations over a 135 day period 
from 2012 November 8 (JD 2456240) to 2013 March 23 (JD 2456375). 
Since solar time drifts with respect to local sidereal time (LST), this observing campaign
yielded more samples of certain LSTs (and hence, sky positions) than others. 
For the power spectrum analysis, we use observations between 0:00 and 8:30 hours
LST.  This range corresponds to
a ``cold patch" of sky away from the galactic center where galactic synchrotron power is minimal,
but also accounts for the weighting of coverage in LST.
Figure \ref{fig:coverage} shows our observing field with the contours labeling
the beam weighted observing time relative to the peak, directly over head the
array.

The PAPER-64 correlator processes a 100--200 MHz bandwidth, first
channelizing the band into 1024 channels of width 97.6 kHz, and then
cross multiplying every antenna and polarization with one another for a total of
8256 cross products, including auto correlations.  Following the architecture 
in \citet{parsons_et_al2008}, this
correlator is based on CASPER\footnote{\url{http://casper.berkeley.edu}} open-source
hardware and signal processing libraries \citep{parsons_et_al2006}.  
Sixteen ROACH boards each hosting eight 8-bit analog-to-digital
converters digitize and channelize antenna inputs. New to this correlator relative to previous PAPER correlators \citep{parsons_et_al2010},
the cross multiplication engine is implemented on eight servers each receiving
channelized data over two 10 Gb Ethernet links.  Each server hosts
two NVIDIA GeForce 580 GPUs running the open-source cross-correlation code developed
by \citet{clark_et_al2013}.
Visibilities are integrated for 10.7 s on the GPUs before
being written to disk.  All polarization cross-products are saved, although the
work presented here only made use of the XX and YY polarization products.

\begin{figure*}
\includegraphics[width=2\columnwidth,trim=0cm 7cm 5cm 2cm,clip]{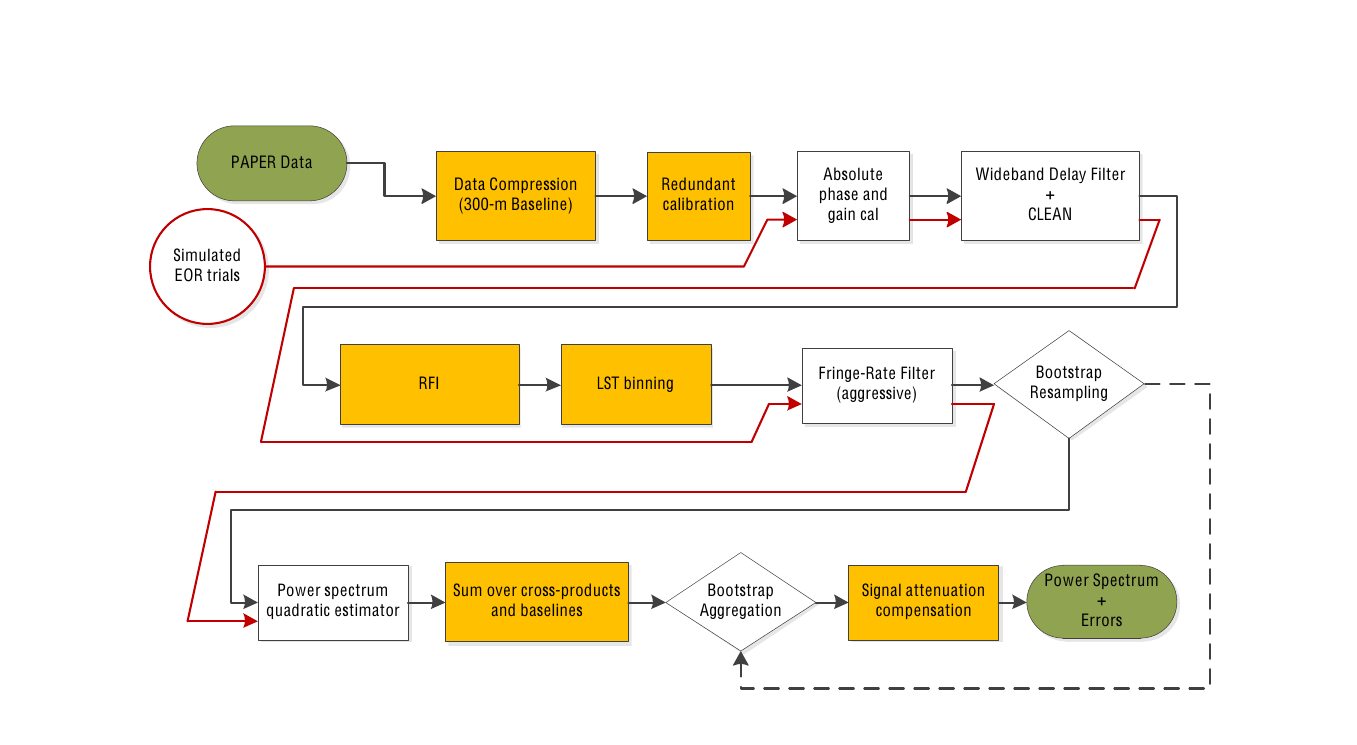}
\caption
{
The stages of power-spectrum analysis. Black lines indicate data flow; red lines indicate
Monte Carlo simulations used to 
measure signal loss. Yellow boxes indicate stages that by construction have negligible signal loss.
Signal loss in other stages is tabluted in Table \ref{tbl:sigloss}.
}
\label{fig:flowchart}
\end{figure*}


\section{Calibration}\label{sec:calib}

Foreground contamination and signal sensitivity represent the two major concerns for $21\,\textrm{cm}$
experiments targeting power spectrum measurements. Sources of foregrounds include
galactic synchrotron radiation, supernova remnants, and extragalactic radio sources.
In the low-frequency radio band (50--200 MHz) where $21\,\textrm{cm}$ reionization
experiments operate, emission from these foregrounds is brighter than the
predicted reionization signal by several orders of magnitude
\citep{santos_et_al2005,ali_et_al2008,deoliveira2008,jelic_et_al2008,bernardi_et_al2009,bernardi_et_al2010,ghosh_et_al2011}.
However, the brightest foregrounds are spectrally smooth, and this provides an
important hook for their isolation and removal
\citep{liu_et_al2009,petrovic_oh2011,liu_tegmark2012}.  Unfortunately,
interferometers, which are inherently chromatic
instruments, interact with spectrally smooth foregrounds to produce unsmooth features that
imitate line of sight Fourier modes over cosmological volumes (P12b; \citealt{morales_et_al2006a,bowman_et_al2009}).
One approach to solving this problem involves an ambitious calibration and modeling approach to spatially localize and
remove foreground contaminants 
\citep{liu_et_al2008,bowman_et_al2008,harker_et_al2009,sullivan_et_al2012,chapman_et_al2013}.
Perhaps the most impressive example of this approach is being undertaken by LOFAR, where dynamic ranges of 4.7 
orders of magnitude have
been achieved in synthesis images \citep{yatawatta_et_al2013}, although it is expected that additional
suppression of smooth-spectrum foreground emission will be necessary \citep{chapman_et_al2013}.

The analysis for this paper employs a contrasting
approach based on the fact that the chromaticity of an interferometer
is fundamentally related to the length of an interferometric baseline.  This relationship, known
colloquially as ``the wedge," was 
derived analytically (P12b; \citealt{vedantham_et_al2012,nithya_et_al2013,liu_et_al2014a,liu_et_al2014b}), and has been confirmed in 
simulations \citep{datta_et_al2010,hazelton_et_al2013} and observationally
\citep{pober_et_al2013,dillon_et_al2013b}.  As described in P12b, the wedge is the result of the delay
between when a wavefront originating from foreground emission
arrives at the two antennas in a baseline.  The fact that this delay is bounded by the light-crossing
time between two antennas (which we call the ``horizon limit'' since such a wavefront would have to 
originate from the horizon) places a fundamental bound on the chromaticity of
an interferometric baseline.  So far, PAPER has had the most success in exploiting this bound (P14; \citealt{jacobs_et_al2014}). 
In this analysis, we continue to use the properties of the 
wedge in order to isolate and remove smooth
spectrum foregrounds.

As illustrated in Figure \ref{fig:flowchart},
our analysis pipeline begins by running a compression
algorithm to reduce the volume of our raw data by a factor of 70.
As described in Appendix A of P14, this is achieved by first performing statistical flagging to remove
radio frequency interference (RFI) at the 6$\sigma$ level, applying low-pass delay and fringe-rate filters that limit signal variation
to delay scales of $|\tau|\la1 \mu$s and fringe-rate scales of $f\la23$ mHz, and then 
decimating to critical Nyquist sampling rates of 493 kHz along the frequency axis
and 42.9 s along the time axis.  We remind the reader that while information is lost in this compression,
these sampling scales preserve emission between
$-0.5\le k_\parallel\le 0.5\hMpci$ that rotates with the sky, making this an essentially lossless compression 
for measurements of the $21\,\textrm{cm}$ reionization signal in these ranges.

After compression, we calibrate in two stages, as described in more detail below.  
The first stage (Section \ref{sec:relcal}) uses instantaneous redundancy to solve for the majority of the 
per-antenna internal dof in the array.  In the second stage (Section \ref{sec:abscal}), standard self-calibration is used 
to solve for a smaller number of
absolute phase and gain parameters that cannot be solved by redundancy alone. 
After suppressing foregrounds with a
wide-band delay filter (Section \ref{sec:wbd_filtering}) and additional RFI flagging and crosstalk removal, 
we average the data in LST (Section \ref{sec:lstbin}) and apply a
fringe-rate filter (Section \ref{sec:frf}) to combine time-domain data. 
Finally, we use an
OQE (Section \ref{sec:oqe}) to make our estimate of the $21\,\textrm{cm}$ power spectrum.

\subsection{Relative Calibration}\label{sec:relcal}

Redundant calibration has gained attention recently as a particularly powerful
way to solve for internal dof in radio interferometric measurements without simultaneously
having to solve for the distribution of sky brightness 
(\citealt{wieringa1992,liu_et_al2010,noorishad_et_al2012,marthi_chengalur2014,zheng_et_al2014}; P14).
The grid-based configuration of PAPER antennas allows a large number of antenna
calibration parameters to be solved for on the basis of redundancy (P14; P12a; 
\citealt{zheng_et_al2014}).  Multiple baselines of the same length and
orientation measure the same sky signal. Differences between redundant
baselines result from differences in the signal chain, including amplitude and
phase effects attributable to antennas, cables, and receivers.  Redundant
calibration only constrains the relative complex gains between antennas and is
independent of the sky. Since redundant calibration preserves signals common to
all redundant baselines, this type of calibration does not result in signal loss.


In practice, redundant calibration often takes on two flavors: log calibration (LOGCAL) and
linear calibration (LINCAL) \citep{liu_et_al2010,zheng_et_al2014}. LOGCAL uses 
logarithms applied to visibilities, 
\begin{equation}
    v_{ij} = g_{i}^{\ast}g_{j}y_{i-j} + n_{ij}^{res},
\end{equation}
where $g$ denotes the complex gain of antennas indexed by $i$ and $j$, and $y$
represents the ``true" visibility measured by the baseline, to give
a linearized system of equations
\begin{equation}\label{eqn:logcal}
    \log{v_{ij}} = \log{g_{i}^{*}} + \log{g_{j}} + \log{y_{i-j}},
\end{equation}
In solving for per-antenna gain parameters with
a number of measurements that scales quadratically with antenna number, redundancy gives 
an over-constrained
system of equations that can be solved
using traditional linear algebra techniques.
While LOGCAL is useful for arriving at a coarse solution from initial estimates that are far
from the true value, LOGCAL has the shortcoming of being a biased by the asymmetric behavior
of additive noise in the logarithm \citep{liu_et_al2010}.

LINCAL, on the other hand, uses a Taylor expansion of the visibility around initial
estimates of the gains and visibilities, 
\begin{equation}\label{eqn:lincal}
v_{ij} = g_{i}^{0*}g_{j}^{0}y_{i-j}^{0} + g_{i}^{1*}g_{j}^{0}y_{i-j}^{0} +
         g_{i}^{0*}g_{j}^{1}y_{i-j}^{0}+g_{i}^{0*}g_{j}^{0}y_{i-j}^{1},
\end{equation}
where $0$ denotes initial estimates and $1$ represents the perturbation to the
original estimate and is the solutions we fit for.  Using initial estimates
taken from LOGCAL, LINCAL constructs an unbiased estimator.

%
%

\begin{figure*}
\centering
\includegraphics[width=1.5\columnwidth]{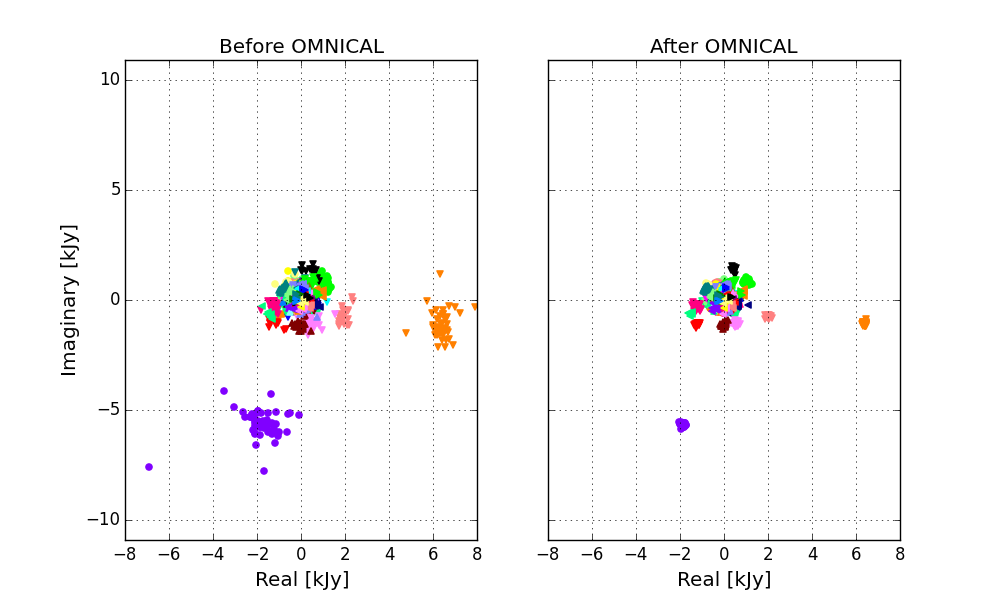}
\caption{
PAPER visibilities plotted in the complex plane before (left) and
after (right) the 
application of the improved redundancy-based calibration with OMNICAL
\citep{zheng_et_al2014}.  All baselines in the array measured at 159 MHz for a
single time integration are plotted.  Instantaneously redundant baselines are
assigned the same symbol/color.  The tighter clustering of redundant
measurements with OMNICAL indicates improved calibration.
} 
\label{fig:omniview}
\end{figure*}

Redundant calibration was performed using 
OMNICAL\footnote{https://github.com/jeffzhen/omnical} --- an open-source
redundant calibration package that is relatively instrument agnostic
\citep{zheng_et_al2014}. This package implements both LOGCAL
and LINCAL, solving for a complex gain solution per antenna, frequency, and
integration. The solutions are then applied to visibilities and the results are
shown in Figure \ref{fig:omniview}.

\begin{figure}
\centering
\includegraphics[width=\columnwidth]{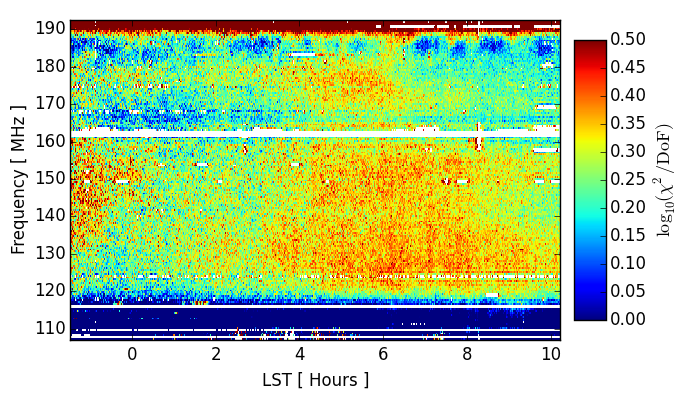}
\caption{
Log of $\chi^{2}$ per degree of freedom of all baseline residuals after the application of OMNICAL.
The plot comprises a observations over one day, with a frequency resolution of
493 kHz and a time resolution of 42.9 s.
} \label{fig:chi2}
\end{figure}

In addition to solving for gain solutions, OMNICAL also characterizes the
quality of the calibration parameters by calculating the $\chi^{2}$ for every
integration. As defined in \citet{zheng_et_al2014}, 
\begin{equation}\label{eqn:chi2}
    \chi^{2} = \sum_{ij}\frac{|v_{ij} - y_{i-j}g^{\ast}_{i}g_{j}|^{2}}{\sigma^{2}_{ij}},
\end{equation}
where $\sigma^{2}$ is the noise in the visibilities. The $\chi^{2}$
measures sum of the deviation of measured visibilities to that of the best fit model
derived from the LINCAL relative to a noise model, and gives us a tool to use in order to check the
quality of our data. The number of dof, as defined in \citealt{zheng_et_al2014}, is given by 
\begin{align}
    \text{dof} &= N_{\text{measurements}} - N_{\text{parameters}} \\\notag
               &= 2N_{\text{baselines}} - 2(N_{\text{antennas}} + N_{\text{unique baselines}}),
\end{align} 
and is effectively the number of visibilities for which
$\chi^{2}$ is calculated. If the data are noise-dominated, 
$\chi^{2}/\text{dof}$ is drawn from a $\chi^{2}$ distribution with $\mu=1$ and
$\sigma^{2} = 2/\text{dof}$. The calculated $\chi^{2}/\text{dof}$ for every
frequency and integration of a fiducial day of observation in this season and
for the fiducial power spectrum baselines is shown in Figure \ref{fig:chi2},
demonstrating the stability of the PAPER instrument.

We measure a mean $\chi^{2}/\text{dof}$ of 1.9.  This
indicates that the redundant calibration solutions, while a substantial improvement
over the previous PAPER-32 calibration (P14), do not quite result in residuals that are thermal noise dominated.
Possible sources of this excess include instrumental crosstalk and poorly performing signal chains.
While the latter will be down-weighted by the inverse of the estimated signal covariance described
in Section \ref{sec:oqe}, crosstalk is a defect in the data that must be addressed.
Crosstalk caused by the cross-coupling of signals between antennas
reveals itself as a static complex bias to a
visibility that varies on timescales much longer than typical fringe rates.
This effect 
skews the distribution of the $\chi^2$ of the residuals away from 1.
To minimize crosstalk, we first use OMNICAL to solve for antenna-dependent gains,
and then average the residual deviations from redundancy
over 10 minute windows before subtracting
the average from the original visibilities. This
crosstalk removal preserves signals common to redundant baseline groups (such as the $21\,\textrm{cm}$ signal).
Unfortunately, it also preserves a term that is the average of the crosstalk of all baselines
in the redundant group.  This residual crosstalk is removed by a fringe-rate filter later
in the analysis.

\subsection{Absolute Calibration}\label{sec:abscal}
%

\begin{figure}
\centering
\includegraphics[width=\columnwidth]{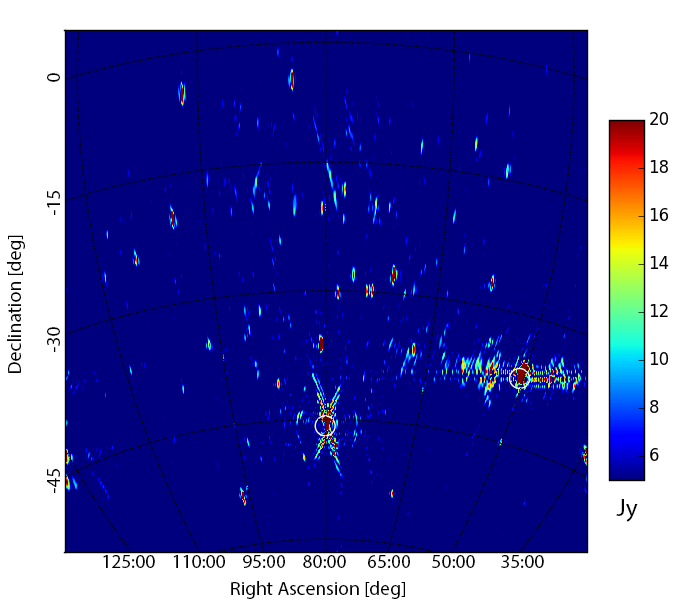}
\caption{
PAPER-64 image of a field including Pictor A and Fornax A, with white circles
indicating catalog positions \citep{jacobs_et_al2011}. Image was synthesized with two hours
of visibilities while Pictor A was in transit and 53 MHz of instantaneous
bandwidth from 120 to 173 MHz.  Image quality is limited by the redundant
configuration of the array (e.g. grating lobes as a result of periodic antenna
spacing, elongated lobes arising from poor uv-coverage in the north-south
direction).  Nonetheless, this image demonstrates accurate phase calibration
over a wide field of view.
} \label{fig:field_image}
\end{figure}

After solving for the relative complex gains of the antennas using redundant
calibration, an overall phase and gain calibration remains unknown. We use the
standard self calibration method for radio interferometers to solve for the
absolute phase calibration. We used Pictor A, Fornax A, and the Crab Nebula to
fit for the overall phase solutions. Figure \ref{fig:field_image} shows an image
of the field with Pictor A (5:19:49.70, -45:46:45.0)  and Fornax A
(3:22:41.70,-37:12:30.0).

%
%
%

We then set our over all flux scale by using Pictor A as our calibrator source
with source spectra derived in \citet{jacobs_et_al2013}, 
\begin{equation}
    S_{\nu} = S_{150}\times\left(\frac{\nu}{150MHz}\right)^{\alpha},
\end{equation}
where $S_{150} = 381.88~\text{Jy} \pm 5.36$ and $\alpha = -0.76 \pm 0.01$, with
1$\sigma$ error bars.

\begin{figure}
\centering
\includegraphics[width=\columnwidth]{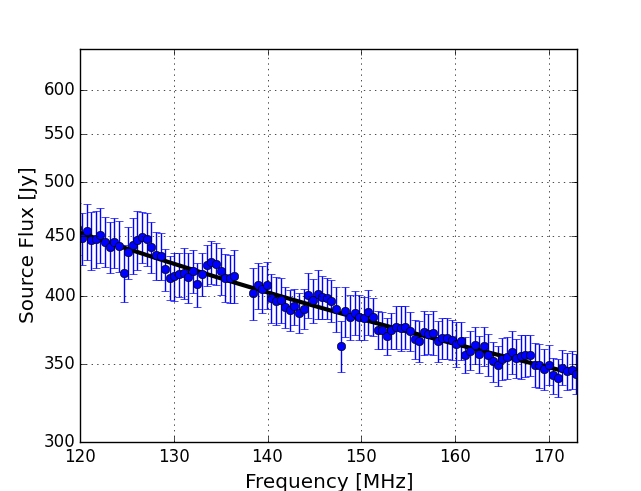}
\caption{
Measured spectrum of Pictor A in Stokes I (blue) relative to its catalog
value (black; \citealt{jacobs_et_al2013}).  Flux density measurements are
extracted from images of Pictor A, made independently for each frequency channel in
10 minutes snapshots as Pictor transits between hour angles of -1:49
and 1:10.  Each measurement is then divided by the PAPER beam model and
averaged to obtain the measured spectrum, which serves to characterize the flux
scale of the PAPER-64 observations. Error bars indicate 68\% confidence
intervals, derived from the Gaussian fits in the source extractor used to
measure the flux density in PyBDSM, combined from all snapshots.
}\label{fig:pic_spec}
\end{figure}

To derive the source spectrum from our measurements, we use data that have been
LST-averaged prior to the wide-band delay filter described in Section
\ref{sec:wbd_filtering}, for the hour before and after the transit of Pictor A.
We image a $30^\circ \times 30^\circ$ field of view for every frequency channel
for each 10 minute snapshot and apply uniform weights to the gridded
visibilities. We account for the required three-dimensional Fourier transform in
wide field imaging by using the w-stacking algorithm implemented in WSclean
\citep{offringa_et_al2014} – although we note that the standard w-projection
algorithm implemented in CASA\footnote{http://casa.nrao.edu} gives similar
performances as the PAPER array is essentially instantaneously coplanar.  A
source spectrum is derived for each snapshot by fitting a two-dimensional
Gaussian to Pictor A by using the
PyBDSM\footnote{http://www.lofar.org/wiki/doku.php?id=public:user\_software:pybdsm}
source extractor. Spectra are optimally averaged together by weighting them with
the primary beam model evaluated in the direction of Pictor A. To fit our
bandpass, we divide the model spectrum by the measured one and fit a 9th order
polynomial over the 120-170 MHz frequency range. Figure \ref{fig:pic_spec} shows
the calibrated Pictor A spectrum and the model spectrum from
\citet{jacobs_et_al2013}. Also plotted are the $1\sigma$ error bars derived from the PyBDSM source extractor and averaged over the multiple snapshots used after being weighted by the beam-squared.

Fitting a polynomial to the bandpass has the potential for signal loss which
would include suppressing modes that may contain the cosmological signal. In order to
quantify the signal loss associated with fitting a ninth degree polynomial to
the bandpass, we run a Monte Carlo simulation of the effect the bandpass has on
a model 21-cm reionization signal. We construct a model baseline visibility as a Gaussian
random signal 
multiplied by the derived bandpass for every independent mode measured. We
calculate the total number of independent modes by counting the number of
independent uv-modes sampled for the different baseline types over the two hour
time interval used to measure the bandpass. We average each mode together and
fit a 9th degree polynomial. Using this as our measured bandpass for this
simulated signal, we finally compare the power spectrum from the output of the
simulated signal to the input power spectrum as a function fo $k$-mode.  We
find that between $-0.06 < k < 0.06$, the width of our wideband delay filter
described below, the signal loss is less than $3\%$ and at the mode right
outside the above limit is $2\times{10^{-7}}\%$. We apply the latter correction
factor for all modes outside the width of the delay filter to the final power
spectrum.

\subsection{Wideband Delay Filtering}\label{sec:wbd_filtering}

Before implementing our foreground removal techniques, we combine the two
linear polarizations for an estimate of Stokes I as per \citet{moore_et_al2013}.
Namely, Stokes I can be estimated as 
\begin{equation}\label{eqn:stokesi}
    V_{\rm I} = \frac12(V_{\rm XX}+V_{\rm YY}),
\end{equation}
where $V_{\rm XX}$ and $V_{\rm YY}$ are the visibilities of the two linear
polarizations measured by the interferometer. There are some important caveats
to the estimate of Stokes I provided by Equation \eqref{eqn:stokesi}. One
important caveat is that it neglects the beam asymmetry  between the two linear
polarization states. This mismatch can cause polarization leakage from Stokes
Q into Stokes I, thus contaminating  our measurement of the power spectrum with any polarized emission from the sky.
This effect for PAPER, as shown in \citet{moore_et_al2013}, leaks 4\% of Q in to
I  in amplitude ($2.2\times10^{-3}$ in the respective power spectra).  We take the conservative approach and do not correct for this effect, noting that the leakage of Q in to I will result in positive power, increasing our limits.

Foreground removal techniques discussed in the literature include spectral
polynomial fitting \citep{wang_et_al2006,bowman_et_al2009,liu_et_al2009},
principal component analysis
\citep{paciga_et_al2011,liu_tegmark2011,paciga_et_al2013,masui_et_al2013},
non-parametric subtractions
\citep{harker_et_al2009,chapman_et_al2013}, and inverse
covariance weighting
\citep{liu_tegmark2011,dillon_et_al2013a,dillon_et_al2013b,liu_et_al2014a,liu_et_al2014b}, Fourier-mode filtering \citet{petrovic_oh2011}, and per-baseline delay filtering described in
P12b.  This delay-spectrum filtering technique is
well-suited to the maximum redundancy PAPER configuration which is not
optimized for the other approaches where high fidelity imaging is a
prerequisite.   The delay-spectrum foreground filtering method is described in
detail by P14; its application is unchanged here.  In summary; we Fourier
transform each baseline spectrum into the delay domain

\begin{align}\label{eqn:delay_transform}
\tilde{V}_\tau &= \int{W_\nu A_\nu I_\nu
                   e^{-2\pi{i}\tau_{g}}\cdot e^{2\pi{i}\tau\nu}d\nu} \\\notag
                &= \tilde{W}_\tau \ast \tilde{A}_\tau \ast
                   \tilde{I}_\tau \ast
                   \delta(\tau_{g} - \tau),
\end{align}
where $A_\nu$ is the frequency dependent antenna response, $W_\nu$ is a sampling function
that includes RFI flagging and a
Blackman-Harris tapering function that minimizes delay-domain scattering 
from RFI flagging, and $I_\nu$ is the source
spectrum.  In the delay domain, a point source appears as a $\delta$-function at
delay $\tau_{g}$, convolved by the Fourier transforms of the
source spectrum, the antenna response, and the
sampling function. We note that the antenna response effectively determines a finite bandpass,
which imposes a lower bound 
of $1/B \approx 10\,\textrm{ns}$ on the width of any delay-domain convolving kernel.
As per
\citet{parsons_backer2009} and P14, we deconvolve the kernel
resulting from $W(\tau)$ using an iterative CLEAN-like procedure
\citep{hogbom1974} restricting CLEAN components to fall within the horizon plus
a 15-ns buffer that includes the bulk of the kernels convolving the $\delta$-function
in Equation \eqref{eqn:delay_transform}.
To remove the smooth spectrum
foreground emission we subtract the CLEAN components from the original
visibility.

\begin{figure*}
\centering
\includegraphics[width=2\columnwidth]{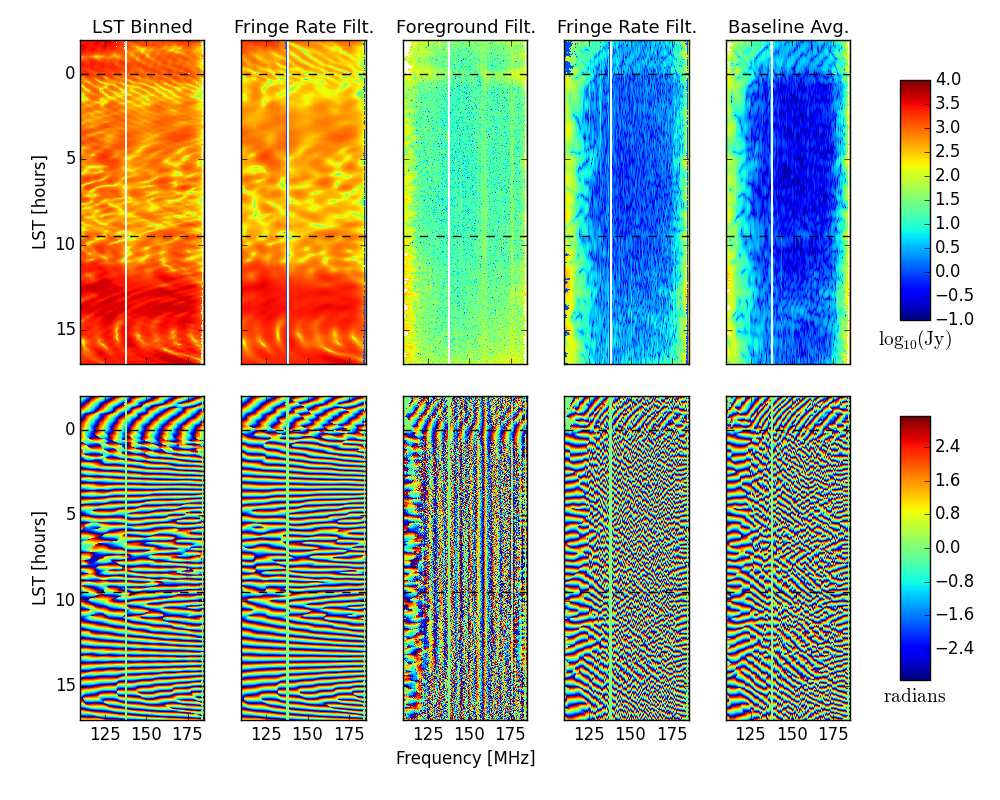}
\caption{
Visibilities measured by a fiducial baseline in the PAPER-64 array, 
averaged over 135 days of observation.  From left to right, columns represent
data that: (1) contain foregrounds prior to the application of a wideband
delay filter or fringe-rate filtering, (2) are fringe-rate filtered but
not delay filtered, (3) are delay filtered at 15 ns beyond the horizon limit but
are not fringe-rate filtered, (4) are both delay and fringe-rate filtered,
and (5) are delay and fringe-rate filtered and have been averaged over all
redundant measurements of this visibility.  The top row shows signal amplitude
on a logarithmic scale; the bottom row illustrates signal phase.
Dashed lines indicate the 0:00--8:30 range in LST used for power spectrum
analysis. The putative crosstalk is evident in the center panel as constant
phase features which do not fringe as the sky.  The two right panels show some
residual signal in the phase structure which is present at low delay. Away from
the edges of the observing band, over four orders of magnitude of foreground
suppression is evident.
} \label{fig:waterfalls}
\end{figure*}

Applying the delay filter to fiducial baselines used in the power spectrum analysis,
foregrounds are suppressed by $\sim$4 orders of magnitude in power, or
 $-40$ dB of foreground suppression, as seen in Figure
\ref{fig:waterfalls}. As discussed in P14, there is a small amount of signal loss
associated with this filter. For the baselines and filter parameters used, the loss was found to be 4.8\% for the
first mode outside of the horizon, 1.3\% for the next mode out, and less than
0.0015\% for the higher modes.  

\subsection{Binning in LST}\label{sec:lstbin}

After the wideband delay filter, we remove a second layer of RFI 
 which was overshadowed by the foreground signal. RFI are excised with a filter
which flags values $3\sigma$ above the median using a variance calculated in a
localized time and frequency window.  

We then average the entire season in LST with 43-s bin widths, matching the
cadence of the compressed data. The full season was 135 days long; of these,
124 days were included in the average. We make two separate LST-binned data
sets, averaging every other Julian day together to obtain an ``even" and ``odd"
dataset. The use of these two data sets allows us to construct an unbiased
power spectrum estimate.


Sporadic RFI events result
in measurements that, in any individual LST bin, deviate from the Gaussian
distribution characteristic of thermal noise.
To catch these events, we compute the median
of a LST bin for each frequency and flag values 3$\sigma$ above the median,
before averaging. 
Since we are narrowing the
distribution of visibilities about the median, the measured thermal noise
variance is not preserved under this filter.  However, since the central
value is preserved, the expectation value of the measured visibility in each
LST bin is unchanged, and there is no associated signal loss for power spectrum
measurements.  Moreover, because errors are estimated empirically
through bootstrapping (see Section \ref{sec:bootstrap}), the slight increase in measurement 
error associated with truncating the tails of the
Gaussian distribution are naturally accounted for.

\subsection{Fringe-rate Filtering}\label{sec:frf}

By averaging visibilities in time, we aim to maximize sensitivity
by coherently combining repeated measurements
of $k$-modes 
before squaring these measurements and averaging over independent $k$-modes
to estimate the power spectrum amplitude.
This is mathematically similar to the more traditional
process of gridding in the $uv$ plane, but applied to a single baseline.
However, rather than applying a traditional box-car average, we can apply a
kernel --- a so-called ``fringe-rate" filter --- that weights different temporal
rates by the antenna beam corresponding to the parts of the sky moving at the
same rate.

For a given baseline and frequency, different parts of the sky
exhibit different fringe-rates.  Maximum fringe rates are found along the
equatorial plane, where the rotation rate of the sky is highest, and zero
fringe rates are found at the poles, where the sky does not rotate and hence
sources do not move through the fringes of a baseline \citep{parsons_backer2009}.
Fringe rates are not constant as a function of latitude. Bins of
constant fringe rate correspond to rings in R.A. and decl., where the east--west
projection of a baseline projected toward a patch of the sky is constant.  We
use this fact in conjunction with the root-mean-squared beam response for each
contour of constant fringe rate to construct a time average kernel or
``fringe-rate filter."

%

As examined in \citet{parsons_et_al2015}, it is possible to tailor fringe-rate filters
to optimally combine time-ordered data for power-spectrum analysis.
Fringe-rate filters can be chosen that
up-weight points of the sky where our instrument is more sensitive and down-weight
those points farther down in the primary beam, which are less sensitive.
For white noise,
all fringe-rate bins will contain the same amount of noise, but the amount of signal
in each bin is determined by the primary beam response on the sky.
By weighting fringe-rate
bins by the rms of the beam response, we can get a net increase in sensitivity.  

Applying this filter effectively weights the data by another factor of the beam
area, changing the effective primary beam response\footnote{The angular area in
Equation \eqref{eqn:delay_pspec} will reflect the new angular area
corresponding to the change in beam area.}, $A(l,m)$ \citep{parsons_et_al2015}.
By utilizing prior knowledge about the beam area, we are selectively
down-weighting areas on the sky contributing little signal. This will result in
a net improvement in sensitivity depending on the shape of the beam and the
decl. of the array. For PAPER, this filter roughly doubles the
sensitivity of our measurements.  

Generally, a fringe-rate filter integrates visibilities in time. For a 
fringe-rate filter, $f_\text{fr}$, the effective integration time can be calculated by comparing the variance statistic before 
and after filtering: 
\begin{equation}\label{eqn:frf_inttime}
    t_{\text{int,after}} = t_{\text{int,before}}\frac{\int{\sigma_{f}^{2}df}}{\int{\sigma_{f}^{2}f_{\text{fr}}^{2}df}},
\end{equation}
where $t_{\text{int,before}}$ is the integration time before filtering, $\sigma_{f}$ denotes the noise variance in fringe rate space and the integral is taken over all possible fringe rates for a given baseline and frequency. As discussed in \citet{parsons_et_al2015}, the
signal re-weighting  associated with this fringe-rate filter can be interpreted as a modification
to the shape of the primary beam.

\begin{figure}\centering
\includegraphics[width=\columnwidth]{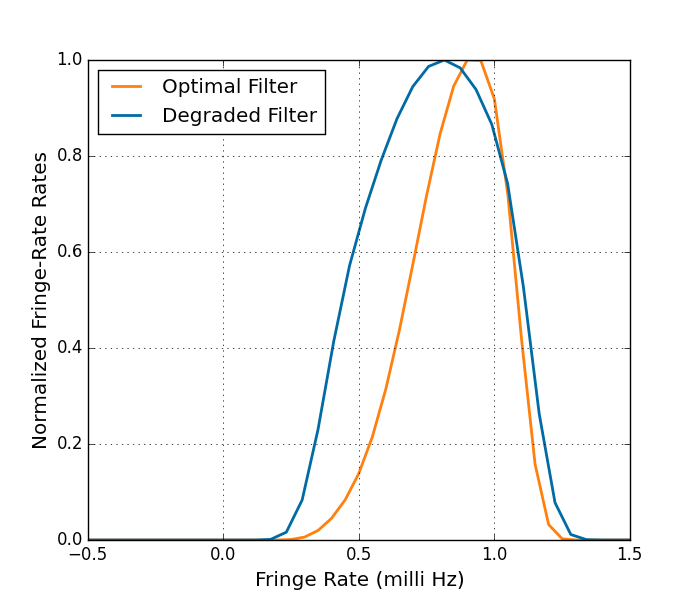}
\caption{
The optimal fringe-rate filter (orange) that and the degraded fringe-rate filter (blue) actually used in the analysis  at 151 MHz, normalized to peak at unity. 
}
\label{fig:fr_preserved_signal}
\end{figure}

For the fiducial baseline at 151 MHz, the integration time, as given in equation \eqref{eqn:frf_inttime}, associated with an optimal fringe rate filter is $3430\, \textrm{s}$. The number of statistically independent samples on the sky decreases from 83 to 1 sample per hour. As discussed in section \ref{sec:sigloss}, empirically estimating a covariance matrix with a small 
number of independent samples can lead to signal loss in the OQE. In order to counteract the signal loss, we degrade the optimal fringe-rate filter, as shown in Figure \ref{fig:fr_preserved_signal}, to have an effective integration time of 1886 s, increasing the number of independent modes to 2 per hour. The fringe rate filter is now sub-optimal, but is still an improvement on the boxcar weighting as used in P14. As documented in Table \ref{tbl:sigloss}, the correction factor for the associated signal loss of the filter we have chosen is 1.39.


We implement the modified filter on a per baseline basis by weighting the
fringe-rate bins on the sky by the RMS of the beam at that same location.
In order to obtain a smooth filter in the fringe-rate domain, we fit a Gaussian
with a hyperbolic tangent tail to this filter. In addition, we multiply this
response with another hyperbolic tangent function that effectively zeros out
fringe rates below $0.2\,\textrm{mHz}$. This removes 
the slowly varying signals that we model as crosstalk. We convolve the
time-domain visibilities with the Fourier transform of the resulting fringe-rate
filter, shown in Figure \ref{fig:fr_preserved_signal}, to produce an averaged
visibility. The effect on the data can be seen in Figure \ref{fig:waterfalls}.

\section{Instrumental Performance}\label{sec:instrument}
\subsection{Instrument Stability}

\begin{figure*}
\centering
\includegraphics[width=2.3\columnwidth]{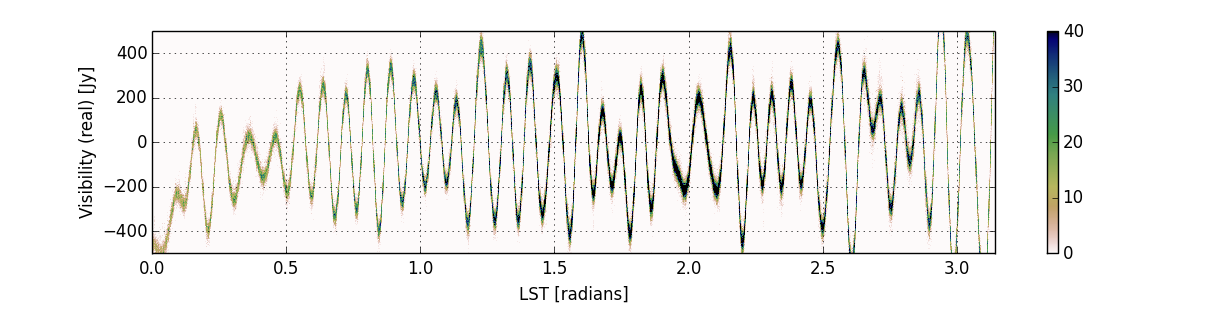}
\caption{Histogram of the real component of all calibrated visibilities
measured over 135 days with every redundant instance of the fiducial baseline at 150
MHz.  Color scale indicates the number of samples falling in an
LST/flux-density bin.  This plot serves to illustrate the stability of the
PAPER instrument and the precision of calibration.  The temporal stability of
a single LST bin over multiple days is shown in Figure \ref{fig:stability}.
}\label{fig:density}
\end{figure*}

\begin{figure}
\centering
\includegraphics[width=\columnwidth]{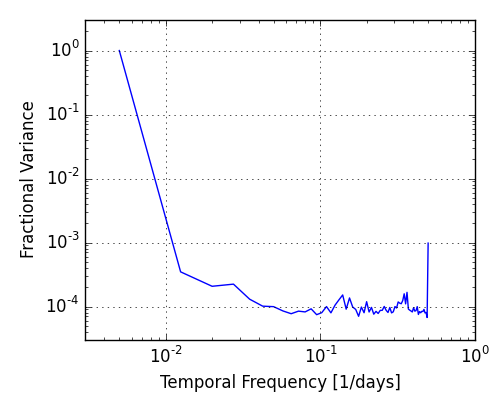}
\caption{
Power spectrum of 135 days of time-series data contributing to a single LST
bin, illustrating the stability of measurements over the observing campaign.
Relative to the average value, variation in the measured value across days (quantified by
variance as a function of time period) is orders of magnitude lower.
The excess at two-day timescales is a beat frequency associated with
the changing alignment of integration windows in the correlator with respect to
sidereal time. 
}\label{fig:stability}
\end{figure}

In order to build sensitivity to the $21\,\textrm{cm}$ reionization signal, it is critical
that PAPER be able to integrate coherently measurements made with different baselines on different days.
Figure \ref{fig:density} shows the visibility repeatability between 
baselines and nights as a function of LST. Specifically, we histogram the real part of
the visibilities for all redundant fiducial baselines 
in a given LST bin for foreground contained data. We see that for a
given LST bin, the spread in values over all the baselines is $\sim$50 Jy which corresponds with our observed
$\Tsys\sim$500K.  We get
more samples per LST bin in the range of 2--10 hr due to our observing
season, therefore the density of points in this LST region is greater, as shown by
the color scale. This density plot shows that redundant baselines agree very well
with one another; OMNICAL has leveled the antenna gains to within the noise.

Delving in a little deeper, we also examine the
stability in time for measurements in a particular LST bin. In order to quantify the stability
in time we extract one channel for a given baseline for every observation day
and LST bin. We then Fourier transform along the time direction for every LST
bin and compute the power spectrum. As shown in Figure \ref{fig:stability},
for time scales greater than one day, we see that signal variance drops by
almost four orders of magnitude, 
with the exception of an 
excess on two-day timescales caused by the changing alignment of the $42.9\,\textrm{s}$
integration timescale relative to a sidereal day.  The implication of this
measurement is that, after calibration, PAPER measurements are sufficiently
stable to be integrated coherently over the entire length of a 135 day
observation. This implies day-to-day stability of better than 1\%, contributing
negligibly to the uncertainties in the data.

\subsection{System Temperature}   

\begin{figure}\centering
\includegraphics[width=\columnwidth]{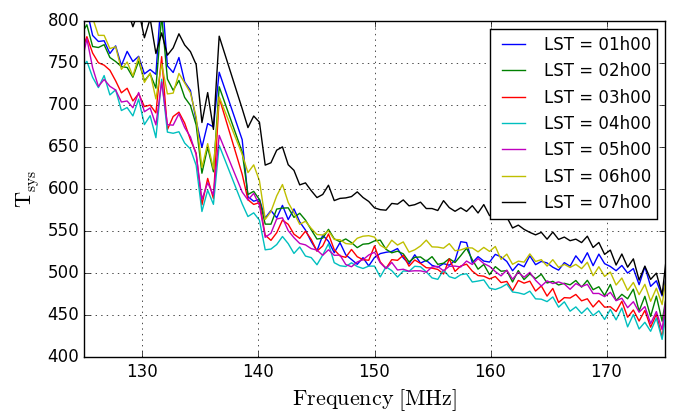}
\caption{System temperature, inferred from the variance of samples falling 
in an LST bin, averaged over one-hour intervals in LST.  The measured value
in the 150--160 MHz range is consistent with previous determinations of
system temperature (\citealt{jacobs_et_al2014}; P14).
}\label{fig:tsys}
\end{figure}

During the LST binning step, the variance of the visibilities that are averaged
together for a given frequency and LST bin are recorded. Using these variances,
we calculate the system temperature as a function of LST, averaging over each
LST hour. 
\begin{equation}
    T_{\rm rms} = \Tsys/\sqrt{2\Delta\nu t}, 
\end{equation}
where $\Delta\nu$ is the bandwidth, $t$ is the integration time, and
$T_{\rm rms}$ is the RMS temperature, or the variance statistic described above.
Figure \ref{fig:tsys} shows the results of this calculation. In this observing
season, the system temperature drops just below previous estimates 
as in P14 and \citet{jacobs_et_al2014} of $\Tsys=560\,\textrm{K}$, at
$\Tsys=500\,\textrm{K}$ at 160 MHz. However, this estimate is more consistent
with the results derived in \citep{moore_et_al2015}, where $\Tsys=505\,\textrm{K}$ at 164
MHz. The change in the system temperature can be attributed to the reduced
range of LST used in the calculation. We note that at 7:00 LST, there is an
increase in the system temperature due to the rising of the galactic plane as
seen in Figure \ref{fig:coverage}.

When calculating the system temperature using the variance in the visibilities
for a given LST and frequency, we take into account the fact that we flag
3$\sigma$ outliers from the median. To calculate an effective correction factor
to account for the filtering, we assume the visibilities follow a Gaussian
distribution which would require a correction factor of 1.34 for the removal of
data points that are 3$\sigma$ above the median. In other words, we are
accounting for the wings of the Gaussian that would contribute to the variance
in the visibility.

Previous estimates
of the system temperature
(P14; \citealt{jacobs_et_al2014}) relied on differencing and averaging
baselines, time samples, and/or frequency channels. The relative agreement
between these various methods of estimating the system temperature provides a
robust measure of the system temperature of the PAPER instrument. Agreement
between the instantaneous measurements of the system temperature, the LST
repetition variance, and the predicted power spectrum noise level (see below)
indicates a robustly stable system with no significant long term instability
contributing appreciable noise.

\section{Power Spectrum Analysis}\label{sec:oqe}

In this section we first
review the OQE formalism, followed by a walk-through of our 
particular applications of the OQE
method to our data. Finally, we discuss the effects of using an empirically estimated 
covariance matrix in our analysis.

%
%
\subsection{Review of OQEs}
We use the OQE method to estimate our power spectrum as
done in \citet{liu_tegmark2011}, \citet{dillon_et_al2013a}, \citet{liu_et_al2014a}, \citet{liu_et_al2014b}, and \citet{trott_et_al2012}.  Here we briefly review the
OQE formalism with an emphasis on our application
to data, which draws strongly from the aforementioned works, but also relies on empirical
techniques similar to those used in P14. The end goal of this analysis is to estimate the $21\,\textrm{cm}$
power spectrum, $P_{21}(\k)$, defined such that 
\begin{equation}
\label{eqn:pspec_def}
    \expval{\widetilde{T}_{b}(\k)\widetilde{T}^{*}_{b}(\k^{\prime})} =
            (2\pi)^{3}\delta^{D}(\k - \k^{\prime})P_{21}(\k),
\end{equation}
where $\widetilde{T}_{b}(\k)$ is the spatial Fourier transform of the brightness temperature
distribution on the sky, $\expval{}$ denotes an ensemble average, and
$\delta^{\text{D}}$ is the Dirac delta function. 

In order to make an estimate of the power spectrum in the OQE formalism, one begins with a data
vector $\x$. This vector could, for example, consist of a list of brightness temperatures on the sky
for an imaging-based data analysis, or (in our case) a list of measured visibilities. We form
the intermediate quantity,
\begin{equation}
\label{eqn:qalpha}
   \hat{q}_{\alpha} = \frac{1}{2}\x^\dagger\mathbf{C}^{-1}\mathbf{Q}_{\alpha}\mathbf{C}^{-1}\x - b_{\alpha},
\end{equation}
which will be needed to form the OQE of our power spectrum.
Here, $\mathbf{C} \equiv \langle \x \x^\dagger \rangle$ is the true covariance matrix of the data vector $\x$, 
$\mathbf{Q}_{\alpha}$ is the operator that takes visibilities into power spectrum
$k$-space and bins into the ${\alpha}$th bin, and $b_{\alpha}$ is the bias to
the estimate that needs to be subtracted off. In general, $\mathbf{Q}_{\alpha}$
represents a family of matrices, one for each $k$ bin indexed by $\alpha$. Each matrix
is defined as
$\mathbf{Q}_{\alpha} \equiv
\frac{\partial{\mathbf{C}}}{\partial p_{\alpha}}$, i.e., the derivative of the covariance
matrix with respect to the band power $p_\alpha$. 
The bandpower $p_\alpha$
can be intuitively thought of as the value of the power spectrum in the $\alpha$th
$k$ bin.  Therefore, $\mathbf{Q}_{\alpha}$ encodes the response of the data
covariance matrix to the $\alpha$th bin of the power spectrum. 

The bias term $b_{\alpha}$ in Equation \eqref{eqn:qalpha} will include contributions
from both instrumental noise and residual foregrounds. Their presence in the data
is simply due to the fact that both contributions have positive \emph{power}. One
approach to dealing with these biases is to model them and to subtract them off,
as is suggested by Equation \eqref{eqn:qalpha}. An alternate approach is to
compute a cross-power spectrum between two data sets that are known to have
the same sky signal but independent instrumental noise realizations. Labeling
these two data sets as $\x_1$ and $\x_2$ and computing
\begin{equation}\label{eqn:qalpha_unbiased}
    \hat{q}_{\alpha} =
\frac{1}{2}\x_{1}^\dagger\mathbf{C}^{-1}\mathbf{Q}_{\alpha}\mathbf{C}^{-1}\x_{2},
\end{equation}
one arrives at a cross-power spectrum that by construction has no noise bias. There is
thus no need to explicitly model and subtract any noise bias, although any residual foreground
bias will remain, since it is a contribution that is sourced by signals on the sky, and therefore
must exist in all our data sets.

The set of $\hat{q}_{\alpha}$s do not yet constitute a properly normalized estimate of
the power spectrum (as evidenced, for example, by the extra factors of $\mathbf{C}^{-1}$). 
To normalize our results, we group the unnormalized bandpowers into a vector $\mathbf{\hat{q}}$
and apply a matrix $\mathbf{M}$ (whose exact form we specify later), so that
\begin{equation}\label{eqn:pspec_norm}
    \mathbf{\hat{p}} = \mathbf{M}\hat{\mathbf{q}}
\end{equation}
is a normalized estimate $\mathbf{\hat{p}}$ of the true power spectrum $\mathbf{p}$. We emphasize
that the vector space that contains $\hat{\mathbf{q}}$ and $\hat{\mathbf{p}}$ is an ``output" vector
space over different $k$-bins, which is separate from the ``input" vector space of the measurements, in which $\x$ and $\mathbf{C}$ reside.

To select an $\mathbf{M}$ matrix that properly normalizes the power spectrum, we must compute
the window function matrix $\mathbf{W}$ for our estimator. The window matrix is defined such that
the true bandpowers $\mathbf{p}$ and our estimates $\hat{\mathbf{p}}$ of them are related by
\begin{equation}\label{eqn:true_pspec_2_est_pspec}
\hat{\mathbf{p}} = \mathbf{W} \mathbf{p},
\end{equation}
so that each row gives the linear combination of the true power that is probed by our estimate. With
a little algebra, one can show that 
\begin{equation}\label{eqn:window_def}
    \mathbf{W} = \mathbf{M}\mathbf{F}, 
\end{equation}
where
\begin{equation}
\label{eq:FisherMatrix}
\mathbf{F}_{\alpha\beta} =
\frac{1}{2}\textrm{tr}(\mathbf{C}^{-1}\mathbf{Q}_{\alpha}\mathbf{C}^{-1}\mathbf{Q}_{\beta}),
\end{equation}
which we have suggestively denoted with the symbol $\mathbf{F}$ to highlight the fact that this turns out to be the Fisher
information matrix of the bandpowers. In order to interpret each bandpower as the weighted
average of the true bandpowers, we require each row of the window function matrix to sum to
unity. As long as $\mathbf{M}$ is chosen in such a way that $\mathbf{W}$ satisfies this criterion, the resulting bandpower estimates
$\hat{\mathbf{p}}$ will be properly normalized.

Beyond the normalization criterion, a data analyst has some freedom over the
precise form of $\mathbf{M}$, which effectively also re-bins the bandpower estimates. One 
popular choice is $\mathbf{M} = \mathbf{F}^{-1}$, which implies that $\mathbf{W} = \mathbf{I}$. Each
window function is then a delta function, such that bandpowers do not
contain leakage from other bins, and contain power from only that bin. However, the disadvantage
of this becomes apparent if one also computes the error bars on the bandpower estimates.
The error bars are obtained by taking the square root of the diagonal of the covariance matrix, which is defined as
\begin{equation}\label{eqn:err_cov}
    \boldsymbol \Sigma = \text{Cov}(\hat{\mathbf{p}}) = \expval{\hat{\mathbf{p}}\hat{\mathbf{p}}^{\dagger}} -
             \expval{\hat{\mathbf{p}}}\expval{\hat{\mathbf{p}}}^{\dagger}.
\end{equation}
Since $\phat = \mathbf{M}\qhat$, it is easily shown that 
\begin{equation}
\label{eq:MFM}
  \boldsymbol  \Sigma = \mathbf{M}\mathbf{F}\mathbf{M}^{\dagger}.
\end{equation}
The choice of $\mathbf{M} = \mathbf{F}^{-1}$ tends to give rather large error bars.
At the other extreme, picking $\mathbf{M}_{\alpha \beta} \propto \delta_{\alpha \beta} / \mathbf{F}_{\alpha \alpha}$ (with the proportionality constant fixed by our normalization criterion) 
leads to the smallest possible error bars \citep{tegmark1997}, at the expense of broader
window functions. In our application of OQEs in the following sections, we will pick an intermediate
choice for $\mathbf{M}$, one that is carefully tailored to avoid the leakage of foreground power
from low $k$ modes to high $k$ modes.

%

\subsection{Application of OQE}
\label{sec:oqe_app}

Here we describe the specifics of our application of the OQE
formalism to measure the power spectrum. Doing so requires defining
various quantities such as $\x$, $\mathbf{C}$, $\mathbf{Q}_\alpha$ for our analysis
pipeline.

First, we consider $\x$, which represents the data in our experiment.  Our data set consists
of visibilities as a function of frequency and time for each baseline in the
array. In our analysis, we group the baselines into three groups of redundant baselines (described in
Section \ref{sec:observations}), 
in the sense that within each group there are multiple copies of the same baseline. In the
description that follows, we first estimate the power spectrum separately for each group.
Power spectrum estimates obtained from the different redundant groups are then combined in a set of averaging and bootstrapping steps described in Section \ref{sec:bootstrap}. Note that because
our data have been fringe-rate filtered in the manner described in Section \ref{sec:frf},
we may reap all the benefits of coherently integrating in time simply by estimating the power spectrum
for every instant in the LST-binned data before averaging over the time-steps within the LST-binned day \citep{parsons_et_al2015}.

For the next portion of our discussion, consider only the data within a single redundant group. Within each group there are not only multiple identical copies of the same baseline, but in addition (as discussed in
Section \ref{sec:wbd_filtering}), our pipeline also constructs two LST-binned data sets, one from binning all even-numbered days in our observations, and the other from all odd-numbered days. Thus, 
we have not a single data vector, but a whole family of them, indexed by baseline ($i$) and odd
versus even days ($r$). Separating the data out into independent subgroups allows one to estimate
cross-power spectra rather than auto-power spectra in order to avoid the noise bias, as discussed in the previous section. The data vectors take the form
\begin{equation}
\label{eqn:xvectdef}
\mathbf{x}^{ri}(t) = \left( \begin{array}{c}
V^{ri} (\nu_1, t) \\
V^{ri} (\nu_2, t) \\
\vdots \\
\end{array}
\right), 
\end{equation}
where $V^{ri} (\nu, t)$ is the visibility at frequency $\nu$ at time $t$. Each data vector is 
20 elements long, being comprised of 20 channels of a visibility spectrum spanning $10\,\textrm{MHz}$ of bandwidth centered on $151.5\,\textrm{MHz}$.
%

Having formed the data vectors, the next step in Equation \eqref{eqn:qalpha} is to
weight the data by their inverse covariance. To do so, we of course require the covariance
matrix $\mathbf{C}$, which by definition, is the ensemble average of
$\x\x^\dagger$, namely $\mathbf{C} = \expval{\x\x^\dagger}$. Unfortunately, in our case the
covariance is difficult to model from first principles, and we must resort to an
empirically estimated $\mathbf{C}$. We make this estimation by taking the time average
of the quantity $\x\x^\dagger$ over 8.5 hr of LST, estimating a different covariance matrix
for each baseline and for odd versus even days. While an empirical determination of the covariance is advantageous
in that it captures features that are difficult to model from first principles, it
carries the risk of cosmological signal loss \citep{switzer_liu2014}. We will discuss and quantify this signal loss
in Section \ref{sec:sigloss}.

\begin{figure*}\centering
\includegraphics[width=2\columnwidth]{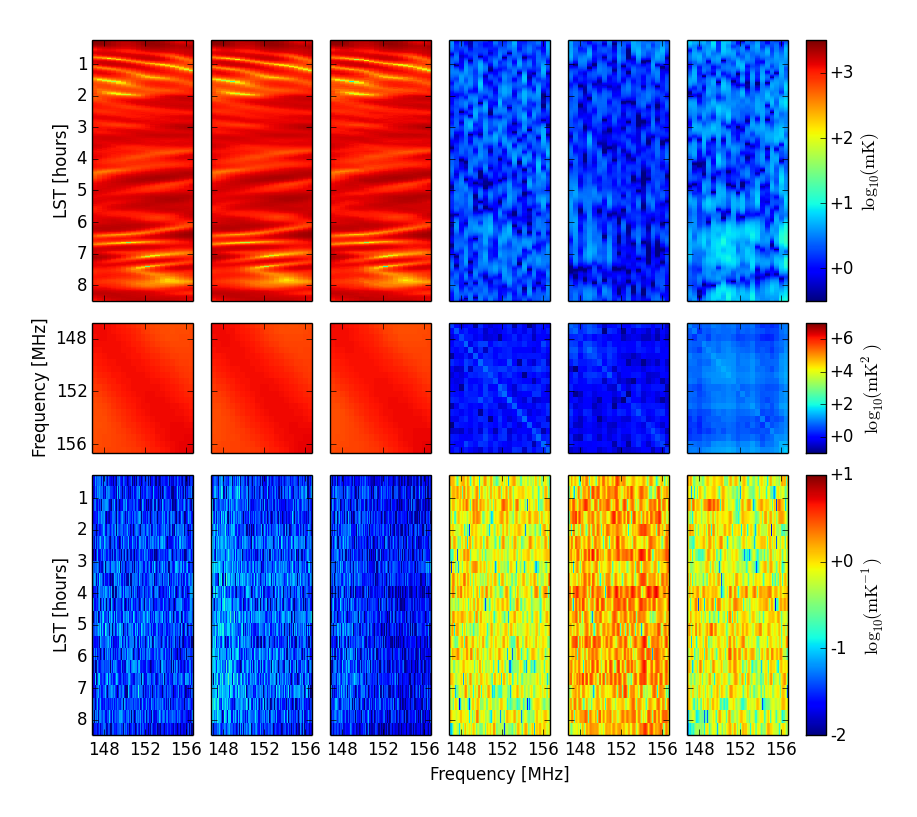}
\caption{
Visibilities before (top row) and after (bottom row) inverse covariance weighting.
Signal covariance (middle row) is estimated empirically, averaging over LST.
The three left/right columns show visibilities from
three different baselines in a redundant group before/after delay filtering, respectively.
} \label{fig:inv_cov}
\end{figure*}

To gain some intuition for the action of $\mathbf{C}^{-1}$ on our data, let us examine the
combination
\begin{equation}\label{eqn:z}
    \mathbf{z}^{ri} =  (\mathbf{C}^{ri})^{-1}\mathbf{x}^{ri}
\end{equation}
for select baselines. This is a crucial
step in the analysis since it suppresses coherent frequency structures (such as those
that might arise from residual foregrounds). Note that the inverse covariance weighting employed
here differs from that in P14, in that P14 modeled and included covariances between
different baselines, whereas in our current treatment we only consider covariances between
different frequency channels. 
Figure \ref{fig:inv_cov} compares the effects of applying
the inverse covariance matrix to a data vector that contains foregrounds (and thus contains
highly correlated frequency structures) to one in which foregrounds have been suppressed
by the wideband
delay filter described in Section \ref{sec:wbd_filtering}. In the figure, the
top row corresponds to the data vector $\mathbf{x}^{ri}$ for three selected baselines in the form
of a waterfall plot of
visibilities, with frequency on the horizontal axis and time on the vertical axis. The
middle section shows the empirical estimate of the covariance by taking the
outer product of $\x$ with itself and averaging over the time axis. Finally,
the last row shows the results of inverse covariance weighting the data,
namely $\mathbf{z}^{ri}$. In every row, the foreground-dominated data are shown
in the left half of the figure, while the foreground-suppressed data are shown in the right half.

\begin{figure}\centering
\includegraphics[width=\columnwidth]{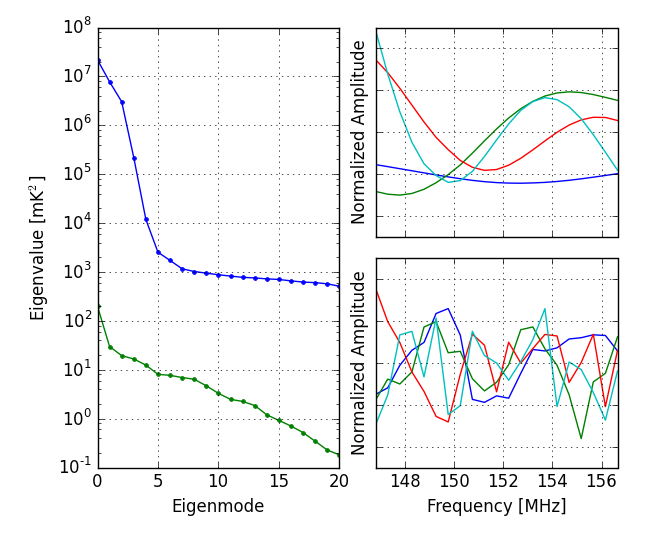}
\caption{
Eigenvalue spectrum of covariance matrices (left) empirically estimated 
from visibilities before (blue) and after (green) delay filtering.
The four strongest eigenmodes of the filtered/unfiltered data are plotted
on the top/bottom panels on the right, respectively.
} \label{fig:eigs}
\end{figure}

Consider the foreground-dominated $\mathbf{x}^{ri}$ in Figure \ref{fig:inv_cov}, and
their corresponding covariance matrices. The strongest modes that are present in the
data are the eigenmodes of the covariance matrix with the largest eigenvalues. Figure
\ref{fig:eigs} shows the full eigenvalue spectrum and the four strongest eigenmodes.
For the foreground-dominated data, one sees that the eigenvalue spectrum is dominated
by the first few modes, and the corresponding eigenmodes are rather smooth, highly suggestive
of smooth spectrum foreground sources. The application of the inverse covariance weighting
down-weights these eigenmodes, revealing waterfall plots in the bottom row of Figure \ref{fig:inv_cov}
that look more noise-dominated. With the foreground-suppressed portion (right half) of Figure \ref{fig:inv_cov},
the initial $\mathbf{x}^{ri}$ vectors already appear noise dominated (which is corroborated by the
relatively noisy form of the eigenvalue spectra in Figure \ref{fig:eigs}). The final $\mathbf{z}^{ri}$ vectors
remain noise-like, although some smooth structure (perhaps from residual foregrounds) has still been removed, and finer scale noise
has been up-weighted.

With intuition established for the behavior of $\mathbf{C}^{-1}$, we may group our identical
baselines into five different sets and average together $\mathbf{z}^{ri}$ vectors for baselines
within the same set. That is, we form
\begin{equation}\label{eqn:presum_oqe}
    \mathbf{z}^{r}_{A} = \sum_{i\in{A}} (\mathbf{C}^{ri})^{-1}\mathbf{x}^{ri},
\end{equation}
where $A$ ranges from $1$ to $5$ and indexes the baseline set. At this point,
we have 10 weighted data vectors $\mathbf{z}$ ($5$ baseline sets, each of which has
an even day and odd day version) for every LST-binned time-step. As discussed in the
previous section, instrumental noise bias may be avoided by forming cross-power spectra
rather than auto-power spectra. Generalizing Equation \eqref{eqn:qalpha_unbiased} to our
present case where we have $10$ different data vectors, we have
\begin{equation}\label{eqn:presum_qalpha}
    \qhat_{\alpha} = \sum_{\substack{A,B,r,s\\r\ne{s},A\ne{B}}}\mathbf{z}^{r\dagger}_{A}\mathbf{Q}_{\alpha}\mathbf{z}^{s}_{B},
\end{equation}
so that auto-power contributions from identical baseline groups or identical even/odd indices
never appear. Residual foreground bias will remain in Equation \eqref{eqn:presum_qalpha}, 
but in order to avoid possible signal loss from an overly aggressive foreground bias removal scheme, 
we conservatively allow the foreground bias to remain. Since foreground power will necessarily
be positive, residual foregrounds will only serve to raise our final upper limits.

In order to implement Equation $\eqref{eqn:presum_qalpha}$, it is necessary to derive a form for
$\mathbf{Q}_\alpha \equiv \partial \mathbf{C} / \partial p_\alpha$. To do so, we follow the delay
spectrum technique of P12a, where it was shown that
\begin{equation}\label{eqn:delay_pspec}
    P(\mathbf{k}_{t\tau}) \approx
\Big(\frac{\lambda^{2}}{2k_{B}}\Big)^{2}\frac{X^{2}Y}{\Omega
B}\expval{\tilde{V}_{i}(t,\tau)\tilde{V}_{j}^{*}(t,\tau)},
\end{equation}
where $V_{i}(t,\tau)$ is the delay transform of baseline visibilities given by Equation \eqref{eqn:delay_transform}, $X$ and
$Y$ are the constants that convert from angles and frequency to the co-moving
coordinate, respectively, $\Omega$ is the power squared beam (see Appendix B of
P14), $B$ is the bandwidth, $\lambda$ is the spectral wavelength, and $k_{B}$ is Boltzmann's constant.
This suggests that in order to estimate the power spectrum from visibilities, one only needs
to Fourier transform along the frequency axis (converting the spectrum into a delay spectrum)
before squaring and multiplying by a scalar. Thus, the role of $\mathbf{Q}_\alpha$ in Equation \eqref{eqn:presum_qalpha} is to perform a frequency Fourier transform on each copy of
$\mathbf{z}$. It is therefore a separable matrix of the form $\Q_{\alpha} =
\mathbf{m}_{\alpha}\mathbf{m}_{\alpha}^{\dagger}$, where $\mathbf{m}_{\alpha}$ is a
complex sinusoid of a specific frequency corresponding to delay mode $\alpha$.
We may thus write
\begin{equation}
    \qhat_{\alpha} =
\sum_{\substack{A,B,r,s\\r\ne{s},A\ne{B}}}\mathbf{z}^{r\dagger}_{A}\mathbf{m}_{\alpha}\mathbf{m}_{\alpha}^{\dagger} \mathbf{z}^{s}_{B}.
\end{equation}
With an explicit form for $\mathbf{Q}_\alpha$, one now also has the necessary ingredients
to compute the Fisher matrix using Equation \eqref{eq:FisherMatrix}.

\begin{figure}\centering
\includegraphics[width=\columnwidth]{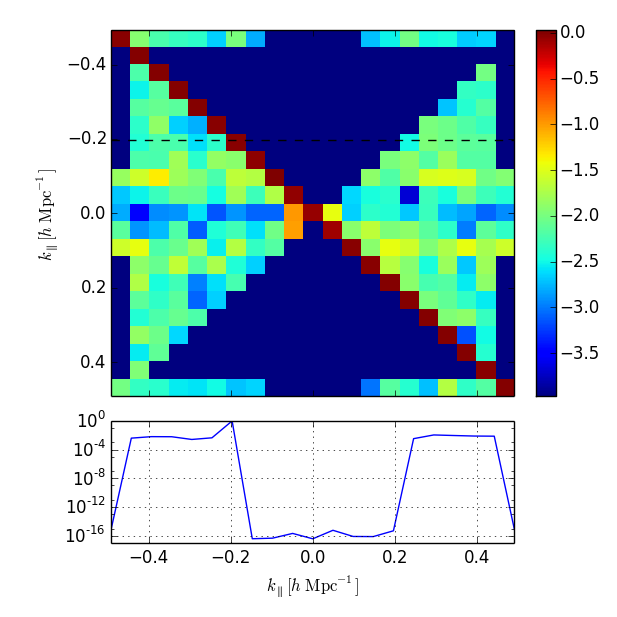}
\caption{
The window function matrix $\mathbf{W}$, as defined in Equation \eqref{eqn:true_pspec_2_est_pspec}.
The $i^{\rm th}$ row corresponds to the window function
used in the estimate of the power spectrum for the $i^{\rm th}$ $k$-mode.
Color scale indicates $\log_{10}\mathbf{W}$.
The inset plot illustrates the window function along the dashed line in the upper panel.
As described in Section \ref{sec:oqe_app}, $\mathbf{M}$ in Equation \eqref{eqn:window_def} has been chosen so that
each window function peaks at the waveband while achieving a high degree of isolation from
at lower $k$-modes that are likely to be biased by foregrounds.
}\label{fig:window_func}
\end{figure}

Having computed the $\qhat_{\alpha}$s, we group our results into a vector $\mathbf{\hat{q}}$.
This vector of unnormalized bandpowers is then normalized to form our final estimates of the
power spectrum $\mathbf{p}$. As noted above, the
normalization occurs by the $\mathbf{M}$ matrix in Equation
\eqref{eqn:pspec_norm}, and can be any matrix of our desire. 
Even though the choices of the normalization matrices described above have certain
good properties, e.g. small error bars or no leakage, we opt for a different
choice of window function, as follows. We first reorder the elements in $\mathbf{\hat{q}}$ (and
therefore in $\mathbf{F}$, $\mathbf{M}$, and $\mathbf{\hat{p}}$ for consistency) so that
the $k$-modes are listed in ascending order, from low $k$ to high $k$, with the exception that we place the highest $k$ bin third after the lowest two $k$ bins. (The reason for this exception will be made apparent shortly). We then take the Cholesky decomposition of
the Fisher matrix, such that $\mathbf{F}=\mathbf{L}\mathbf{L}^{\dagger}$, where
$\mathbf{L}$ is a lower triangular matrix. Following that, we pick ${\mathbf{M}} = \mathbf{D} \mathbf{L}^{-1}$, where $\mathbf{D}$ is a diagonal matrix chosen to adhere to the normalization constraint
that $\mathbf{W} = \mathbf{M} \mathbf{F}$ has rows that sum to unity. In this case, the window function matrix becomes,
$\mathbf{W}=\mathbf{D} \mathbf{L}^{\dagger}$. This means that $\mathbf{W}$ is upper triangular,
and with our ordering scheme, has the consequence of allowing power to leak from high to low $k$,
but not vice versa. Since our $k$ axis is (to a good approximation) proportional to the delay axis, foregrounds preferentially appear at low $k$ because their spectra are smooth. Reducing leakage
from low $k$ to high $k$ thus mitigates leakage
of foregrounds into the cleaner, more noise-dominated regions. Additionally, our placement of the highest $k$ bin as the third element in our reordering of 
$\mathbf{\hat{p}}$ prevents leakage from this edge bin that will contain aliased power. Figure
\ref{fig:window_func} shows the resulting window functions. 

Our choice of normalization matrix also has the attractive property of eliminating error correlations
between bandpower estimates. Using Equation \eqref{eq:MFM}, we have
that \begin{equation} 
 \boldsymbol   \Sigma = \mathbf{D} \mathbf{L}^{-1}\mathbf{L}\mathbf{L}^{\dagger}\mathbf{L}^{-\dagger} \mathbf{D}
           = \mathbf{D}^2.
\end{equation}
The error covariance matrix on the bandpowers is thus diagonal, which implies
that our final data points are uncorrelated with one another. This stands in contrast to the power-spectrum estimator used in P14, where the Blackmann--Harris taper function induced correlated errors
between neighboring data points.

\subsection{Covariance Matrix and Signal Loss}
\label{sec:sigloss}
%

We now discuss some of the subtleties associated with empirically estimating the covariance matrix from
the data. Again, the covariance matrix is defined as the ensemble average of the outer
product of a vector with itself, i.e., 
\begin{equation}
    \C = \expval{\x\x^{\dagger}}, 
\end{equation}
where $\x$ is the data (column) vector used in the analysis. In our analysis,
we do not have \emph{a priori} knowledge of the covariance matrix. and thus we
must resort to empirical estimates \citep{dillon_et_al2015}. As we have alluded to above, we replace
the ensemble average with a time average that runs from 0 to 8:30 LST hours.

Since the OQE method for power spectrum estimation requires the inversion
of $\C$, it is crucial that our empirically estimated covariance be a full rank matrix.
With our data consisting of visibilities over $20$ frequency channels, the covariance
matrix is a $20 \times 20$ matrix. Thus, a necessary condition for our estimate to be
full rank is for there to be at least $20$ independent time samples in our average. As noted in Section \ref{sec:frf} the fringe-rate filter used corresponds to averaging time samples for 31 minutes. Over the LST range
used in this analysis, this corresponds to roughly 20 statistically
independent modes in our data after fringe-rate filtering. We therefore have just enough
samples for our empirical estimate, and in practice, our covariance matrices are invertible
and allow OQE techniques to be implemented.

%

Another potential problem that occurs from empirically estimating covariances is that it
leads to models of the covariance matrix that over-fit the noise. In this
scenario, the covariance matrix tells us that there may be modes in the data
that should be down-weighted, for example, but if the empirical covariance estimates are dominated
by noise, these may just be random fluctuations that need not be down-weighted. Said differently, the weighting of the data by the inverse
covariance is heavily influenced by the noise in the estimate of the covariance
matrix and thus has the ability to down-weight valid high-variance samples. 
Over-fitting the noise in this manner carries with it the possibility of cosmological signal loss.
This seems to contradict the conventionally recognized feature of OQEs as
lossless estimators of the power spectrum \citep{tegmark1997}. However,
the standard proofs of this property assume that statistics such as $\mathbf{C}$
are known \emph{a priori}, which is an assumption that we are violating with our
empirical estimates.
%
%


\begin{figure}
\centering
\includegraphics[width=\columnwidth]{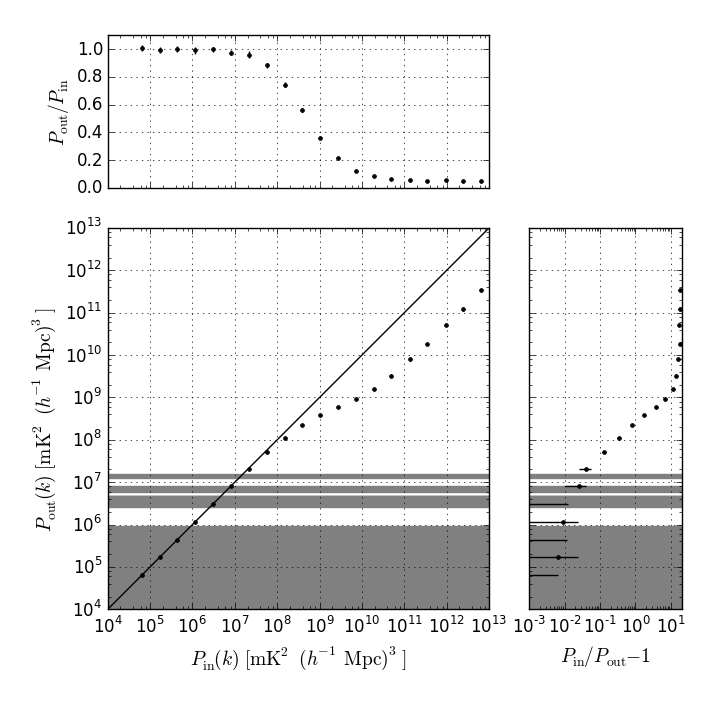}
\caption{
Recovered power spectrum signal as a function of injected signal amplitude.  Shaded regions
indicate the range in measured amplitude of power spectrum modes in Figure \ref{fig:final_pspec}.  
Error bars indicate 95\% confidence intervals as determined from the Monte Carlo simulations
described in Section \ref{sec:sigloss}.
Because
the recovered signal amplitude is a monotonic function of the injected signal amplitude,
it is possible to invert the effects of signal loss in the measured power spectrum values
to infer the true signal amplitude on the sky. Over the range of powers measured, the 
maximum correction factor $P_{\rm in}/P_{\rm out}$ is less than 1.02 at 97.5\% confidence.
The transition to significantly higher correction factors at larger signal amplitudes
occurs as the injected signal
dominates over the foreground modes present in estimates of the data covariance.
}\label{fig:signal_loss}
\end{figure}

In order to deal with possible signal loss, we perform simulations of our analysis pipeline, 
deriving correction factors that must be applied to our final constraints. We simulate visibilities for
Gaussian temperature field with a flat amplitude in $P(k)$ that rotates with the
sky, which is fringe-rate filtered in the same way as the data for our fiducial baselines. This signal is processed through our pipeline, and the output power spectrum compared to the input
power spectrum, for various levels of input signal amplitude.
We repeat this for 40 sky realizations at each signal level.  Figure
\ref{fig:signal_loss} shows the resultant signal loss associated with
estimating the covariance matrix from the data.  Error bars were obtained
through bootstrapping.

As a function of the increasing input amplitude of the simulated power spectra,
we find that the ratio of output power to input power decreases, which we interpret
 as signal loss through the use of our empirical OQE of the power spectrum.  
However, since the transfer function through this analysis is an invertible function,
we can correct for the transfer by using the output value to infer a signal loss
that is then divided out to obtain the original input signal level.  In Figure \ref{fig:signal_loss},
we see that
deviations from unity signal transfer begin at
power spectrum amplitudes of $10^{7} \text{mK}^{2} (h^{-1}\rm
\,\text{Mpc})^{3}$. For the range of output power spectrum amplitudes in our
final estimate of the $21\,\textrm{cm}$ power spectrum (Figure \ref{fig:final_pspec}), we
show that signal loss is $<2\%$ at $95\%$ confidence. 

\begin{table}[htp]
\caption{SIGNAL LOSS VERSUS ANALYSIS STAGE}
\begin{center}
\begin{tabular}{rll}
Analysis Stage & Typical Loss & Maximum Loss \\
\hline
Bandpass Calibration &  $< 2 \times 10^{-7}\%$ & 3.0\% \\
Delay Filtering & $1.5\times10^{-3}\%$ & 4.8\% \\
Fringe-rate Filtering & 28.1\% & 28.1\% \\
Quadratic Estimator & $<2.0\%$ & 89.0\% \\
Median of Modes & 30.7\% & 30.7\% \\
\end{tabular}
\end{center}
\label{tbl:sigloss}
\end{table}%

As shown in Table \ref{tbl:sigloss}, the signal loss we characterize for quadratic
estimation of the power spectrum band powers is tabulated along with the signal
loss associated with each other potentially lossy analysis stage (see Figure \ref{fig:flowchart}).
We correct for the signal loss in each stage by multiplying the final power spectrum results
by the typical loss for each stage, except for modes within the horizon limit and immediately
adjacent to the horizon limit, where we apply the maximum signal loss correction to be conservative.

\subsection{Bootstrapped Averaging and Errors}\label{sec:bootstrap}

\begin{figure*}\centering
\includegraphics[width=1.8\columnwidth]{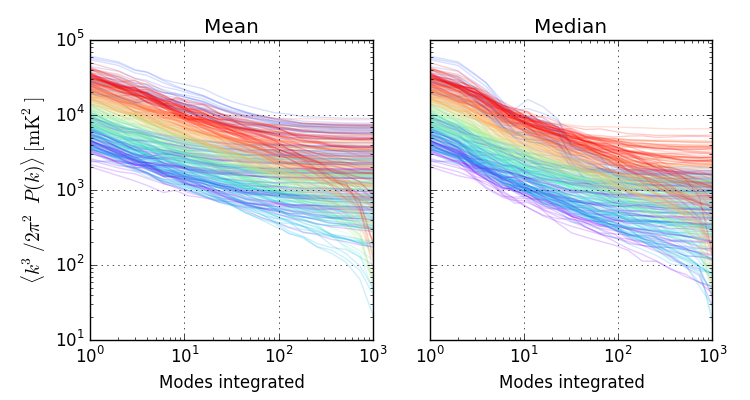}
\caption{
Absolute value of the cumulative mean (left) and median (right), as a function of number of modes 
of the power spectrum band power for
$k_\parallel$ modes ranging from $-0.49$ (red) to $0.44\hMpci$ (violet).
Here, modes are defined as samples from different redundant baseline groups and LSTs.
This Allen variance plot shows modes averaging down as the square root of
number of modes combined until a signal floor is reached.  The difference in
behavior between the mean and median is an indication of outliers
in the distribution of values, likely as a result of foreground contamination.
We use the median in the estimation of the power spectrum in Figure \ref{fig:final_pspec},
along with a correction factor compensating for the difference between the mean and median
in estimating variance.
}\label{fig:pspec_variance}
\end{figure*}

When estimating our power spectra via OQEs, we generate
multiple samples of the power spectrum in order to apply the bootstrap method to
calculate our error bars. In detail, the power spectrum estimation scheme proposed
above requires averaging at several points in the pipeline:
\begin{enumerate}
\item Visibilities are averaged into five baseline groups after inverse covariance weighting (see Equation \eqref{eqn:presum_oqe})
\item Power spectrum estimates from each of the three redundant baseline types (described in Section \ref{sec:observations}) are averaged together.
\item Power spectrum estimates from each LST are averaged together.
\end{enumerate}
With the bootstrapping technique, we do not directly perform these averages. Instead,
one draws random samples within the three-dimensional parameter space specified above,
with replacement, until one has as many random samples as there are total number of parameter
space points. These random samples are then propagated through the power spectrum pipeline
and averaged together as though they were the original data. This forms a single estimate (a ``bootstrap") of $P(\mathbf{k})$. Repeating 
random draws allows one to quantify the inherent scatter---and hence the error bars---in our
estimate of $P(\mathbf{k})$. When plotting $\Delta^2 (k) \equiv k^3 P(k) / 2 \pi^2$ instead of
$P(\mathbf{k})$, we bin power falling in $+k$ and $-k$, and so 
we additionally randomize the inclusion of 
positive and negative $k$ bins.

We compute a total of 400 bootstraps. In combining independent samples for our final power spectrum
estimate, we elect to use the median, rather than the mean, of the samples. One can see the behavior 
of both statistics in Figure \ref{fig:pspec_variance}, where we
show how the absolute value of $\Delta^2(k)$ integrates down as more independent samples are included in the mean and median.
In this plot, one can see modes integrating down 
consistent with a noise-dominated power spectrum until they bottom out on a signal.
In the noise-dominated regime, the mean and the median
behave similarly.  However, we see that the median routinely continues to integrate down as noise for longer.
This is an indication that the mean is skewed by outlier modes, suggesting variations beyond thermal noise. The magnitude of the difference
is also not consistent with the Rayleigh distribution expected of a cosmological power spectrum limited by cosmic
variance.  
For a Rayleigh distribution, the median is $\ln2 \sim 0.69$ times the mean. 
Instead, we interpret the discrepancy as a sign of contributions from foregrounds, which are neither isotropic 
nor Gaussian distributed.  Since median provides 
better rejection of outliers in the distribution that might arise from residual foreground power, we choose to use
the median statistic to combine measurements across multiple modes.
As listed in Table \ref{tbl:sigloss}, we apply a $1/\ln2$ correction factor to our power spectrum estimates to 
infer the mean from the median of a Rayleigh distribution.

\section{Results}\label{sec:results}

\subsection{Power Spectrum Constraints}
To summarize the previous section, we follow 
the power spectrum analysis procedure outlined in Section \ref{sec:oqe_app},
we incoherently combine independent power spectrum measurements made at different
times and with different baseline groups using the median statistic.  As described
in Section \ref{sec:bootstrap}, we bootstrap over all of these independent measurements,
as well as over the selection of baselines included in the power spectrum analysis for
each baseline group, in order to estimate the error bars on the spherically averaged
power spectrum $P(k)$, where positive and negative $k_\parallel$ measurements
are kept separate for diagnostic purposes.  In the estimation of the 
dimensionless power spectrum
$\Delta^{2}(k)\equiv{k^{3}P(k)}/{2\pi^{2}}$, the folding of $\pm k_\parallel$ is
handled along with the rest of the bootstrapping over independent modes.
Finally, the measured values for $P(k)$ and $\Delta^2(k)$ are corrected for signal
loss through all stages of analysis, as summarized in Table \ref{tbl:sigloss}.

\begin{figure*}\centering
\includegraphics[width=2\columnwidth]{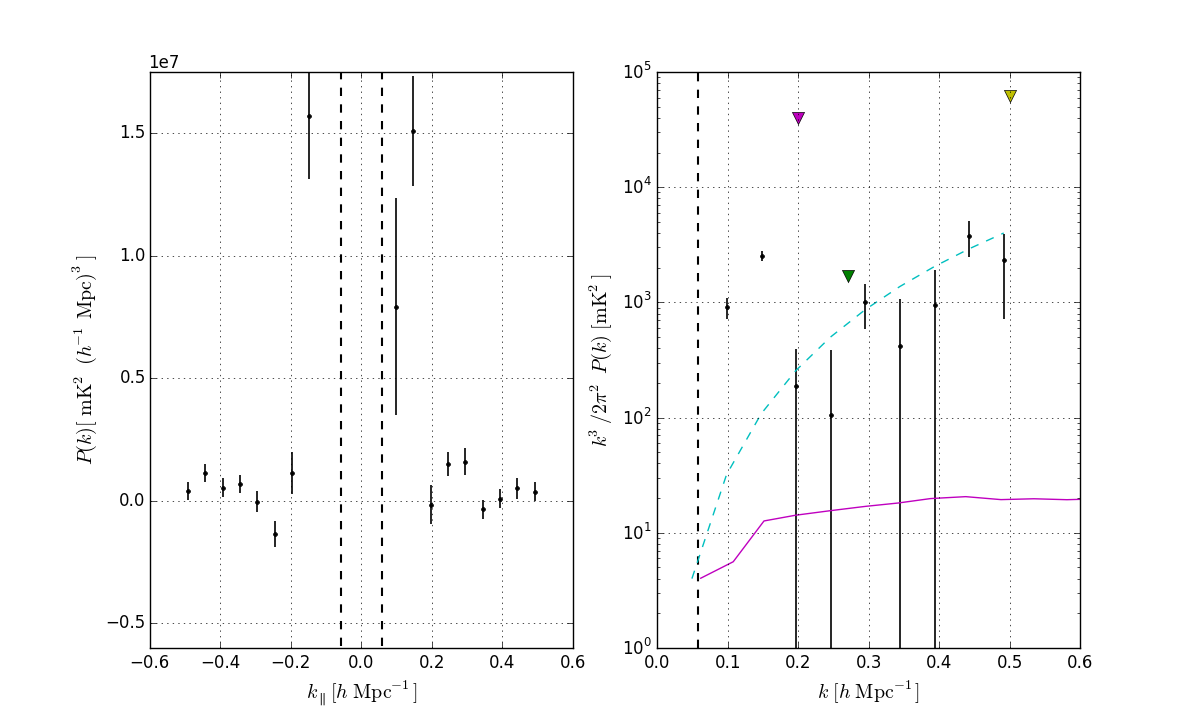}
\caption{
Measured power spectrum (black dots with 2$\sigma$ error bars) at $z=8.4$
resulting from a 135 day observation with PAPER-64.  The dashed vertical lines
at $0.6\hMpci$ show the bounds of the delay filter described in Section
\ref{sec:wbd_filtering}. The predicted 2$\sigma$ upper limit in the absence of the a celestial signal is shown in dashed cyan, assuming $\Tsys=500K$. The triangles indicate 2
$\sigma$ upper limits from GMRT \citep{paciga_et_al2011} (yellow) at $z=8.6$,
MWA \citep{dillon_et_al2013b} at $z=9.5$ (magenta), and the previous PAPER upper
limit (P14) at $z=7.7$ (green). The magenta curve shows a predicted model $21\,\textrm{cm}$ power
spectrum at 50\% ionization \citep{lidz_et_al2008}.
} \label{fig:final_pspec}
\end{figure*}

The final results are plotted in Figure \ref{fig:final_pspec}.
For the first two modes outside of the horizon where $\Delta^2(k)$ is measured, we have
clear detections. We attribute these to foreground leakage from
inside the horizon related to the convolution kernels in Equation \eqref{eqn:delay_transform} (either
from the chromaticity of the antenna response, or from the inherent spectrum of the
foregrounds themselves).  Somewhat more difficult to interpret are the 
$2.4\sigma$ excess at $k\approx0.30\hMpci$ and 
the $2.9\sigma$ excess at $k\approx0.44\hMpci$. Having two such outliers is
unlikely to be chance.

\begin{figure*}\centering
\includegraphics[width=2\columnwidth]{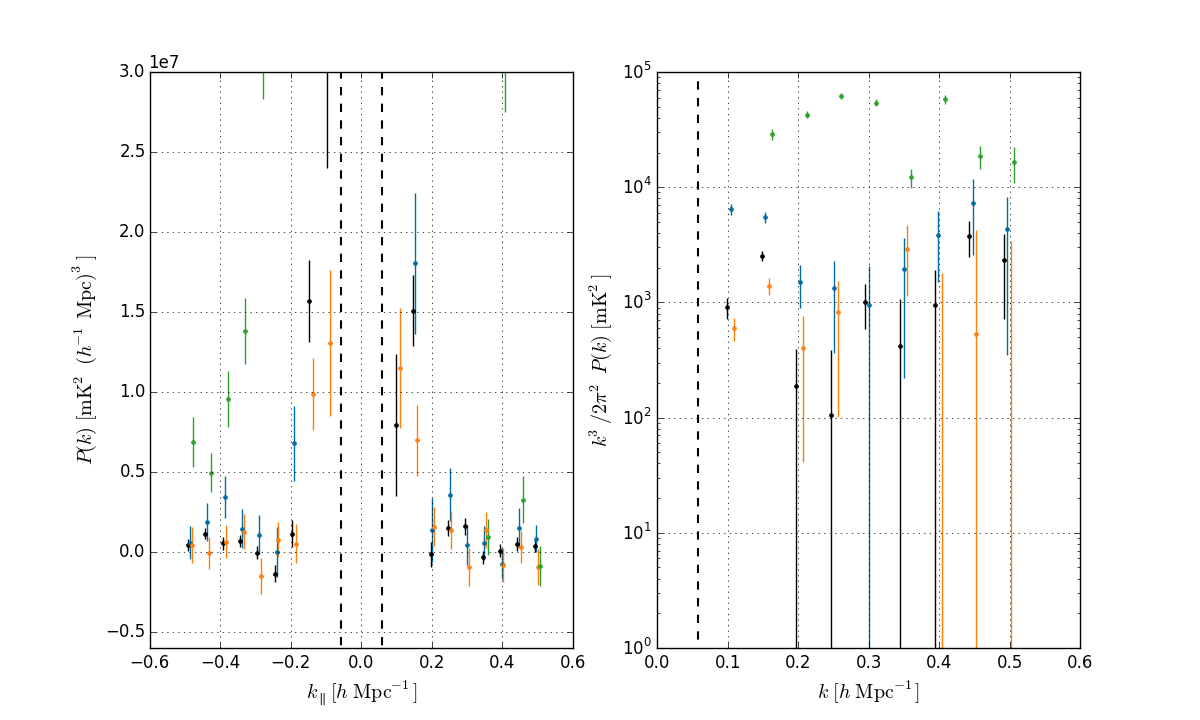}
\caption{
Diagnostic power spectra in the style of Figure \ref{fig:final_pspec}
illustrating the impact of various analysis stages.
The blue power spectrum uses the P14 fringe-rate filter combined with crosstalk removal.
Green illustrates the result using the improved fringe-rate filter, but without crosstalk removal.
A power spectrum derived without the application of OMNICAL is shown in orange.  Black
includes improved fringe-rate filtering, crosstalk removal, and OMNICAL calibration; it is the 
same power spectrum shown in Figure \ref{fig:final_pspec}.
}\label{fig:pspec_comp}
\end{figure*}

In examining the effects on the power
spectrum of omitting various stages of analysis (see Figure \ref{fig:pspec_comp}), we
see a pronounced excess in the green curve corresponding 
to the omission of crosstalk removal in fringe-rate filtering.
While the signal is heavily attenuated in the filtering step, it remains a
possibility that the remaining detections are associated with instrumental crosstalk.
We do note, however, that the qualitative shape of the excess in the crosstalk-removed data
does not appear to match that of the crosstalk-containing data.

Another likely possibility is that the signal might be associated with foregrounds.
Foregrounds, which are not generally isotropically distributed on the sky, are likely
to be affected by the spatial filtering associated with fringe-rate filtering, whereas
a statistically isotropic signal is not.  
Indeed, we see that excesses in many modes
measured with using the P14-stype time-domain filtering (blue in Figure \ref{fig:pspec_comp})
decrease significantly using the improved fringe-rate filter.  
As discussed in \citet{parsons_et_al2015},
the normalization applied to $\Omega_{\rm eff}$ for fringe-rate filtering correctly
compensates for the effect of this filtering on power-spectral measurements
of a statistically isotropic Gaussian sky signal.  We can surmise from any significant change in amplitude of the excess
under fringe-rate filtering that it arises from emission that violates these assumptions.
We conclude, therefore, that this excess is unlikely to be cosmic reionization, and is more
likely the result of non-Gaussian foregrounds.
As discussed earlier, one possible
culprit is polarization leakage \citep{jelic_et_al2010,jelic_et_al2014,moore_et_al2013}, although further
work will be necessary to confirm this.  The interpretation of
the signal as polarization leakage is, however, rather high to be consistent
with recent measurements in Stokes Q presented in \citet{moore_et_al2015},
where the leakage is constrained to be $<$ 100 mK$^{2}$ for all $k$.

That the
excesses at $k\approx0.30$ and 0.44$\hMpci$ are relatively unaffected by the filtering
could be an indication that they are more isotropically distributed, but more likely, it
may mean that the simply arise closer to the center of the primary beam where they are
down-weighted less.
Both excesses appear to be significantly affected by omitting OMNICAL calibration
(orange in Figure \ref{fig:pspec_comp}).  This could be interpreted as indicating the 
excess is a modulation induced by frequency structure
in the calibration solution.  However, OMNICAL is constrained
to prohibit structure common to all baselines, so a more likely interpretation is that 
this faint feature decorrelates without the precision of redundant calibration.  To
determine the nature of these particular excesses, further work will be necessary.

\begin{figure}\centering
\includegraphics[width=\columnwidth]{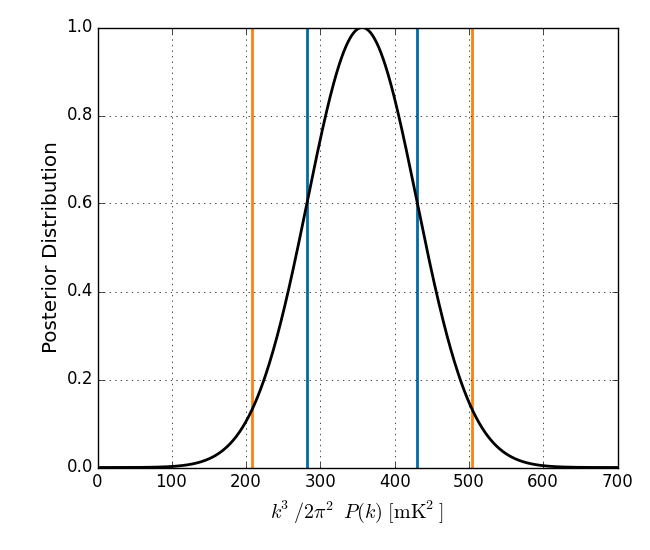}
\caption{
Posterior distribution of power spectrum amplitude for a flat $\Delta^{2}(k)$
power spectrum over $0.15<k<0.5\hMpci$ (solid black),
assuming Gaussian error bars. The blue and orange
vertical lines correspond to the 1$\sigma$ and 2$\sigma$ bounds, respectively.
}
\label{fig:final_posterior}
\end{figure}

In order to aggregate the information presented in the power spectrum into
a single upper limit, we fit a flat $\Delta^2(k)$ model to measurements
in the range $0.15<k<0.5\hMpci$.  We use a uniform prior of amplitudes between
--5000 and 5000 ${\rm mK}^2$, and assume measurement errors are Gaussian.
Figure \ref{fig:final_posterior} shows the posterior distribution of the fit.
From this distribution, we determine a mean of
(18.9 mK)$^2$ and a $2\sigma$ upper limit of \mKlimit.
The measured mean is inconsistent with zero at the 4.7$\sigma$ level, indicating that
we are detecting a clear power spectrum excess at $k>0.15\hMpci$.

We suspect that the excess in our measured power spectrum is likely caused
by crosstalk and foregrounds.  We therefore suggest ignoring the lower bound on the power spectrum amplitude
as not being of relevance for the cosmological signal.  On the other hand, since foreground
power is necessarily positive, the 
2$\sigma$ upper limit of \mKlimit at $z=8.4$, continues to serve as a conservative upper limit. This significantly improves over the previous
best upper limit of $(41~{\rm mK})^2$ at $z=7.7$ reported in P14.
As we show below and in
greater detail in \citet{pober_et_al2015}, this limit begins to have implications
for the heating of the IGM prior to the completion of reionization.

\subsection{Spin Temperature Constraints}

\begin{figure*}\centering
\includegraphics[width=2\columnwidth]{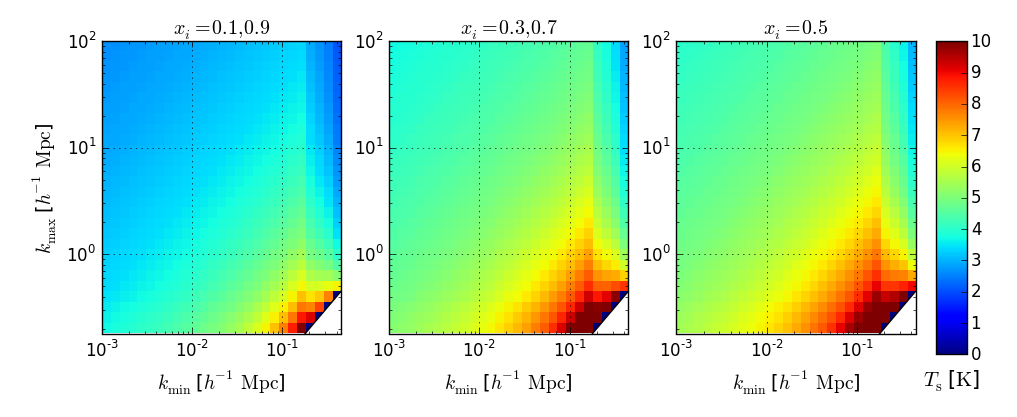}
\caption{Constraints on the 21cm spin temperature at $z=8.4$, 
assuming 
the patchy reionization model in Equations
\eqref{eqn:d2pspec_model} and \eqref{eqn:patchy_bound}, which hold in the limit
that $T_{\rm s}<T_{\rm CMB}$.
} \label{fig:patchy_bound}
\end{figure*}

In this section, we examine the implication of the measured upper limits
on 21cm emission in Figure \ref{fig:final_pspec} on the spin temperature
of the 21cm line at $z=8.4$.
In a forthcoming paper \citep{pober_et_al2015}, we conduct a thorough analysis of the
constraints that can be put on the IGM using a simulation-based framework.
As a complement to that more thorough
analysis, we focus here on a simpler parameterization of the shape
of the 21cm power spectrum signal. 

The brightness temperature of the 21cm signal, $\delta{T_{b}}$, arising from the
contrast between the cosmic
microwave background, $\Tcmb$, and the spin temperature, $\Tspin$, is given
by 
\begin{equation}\label{eqn:tb}
    \delta{T_{b}} = \frac{\Tspin - \Tcmb}{1+z}(1-e^{-\tau})
\approx \frac{\Tspin - \Tcmb}{1+z}\tau,
\end{equation}
where temperatures are implicitly a function of redshift $z$, and
the approximation holds for low optical depth, $\tau$. 
The optical depth is given by \citep{zaldarriaga_et_al2004}
\begin{equation}\label{eqn:tau}
    \tau = \frac{3c^3\hbar A_{10} n_{\text{\tiny{HI}}}}{16k\nu_0^2\Tspin H(z)}
\end{equation}
where $A_{10}$ is the Einstein A coefficient for the 21cm transition,
$n_{\text{HI}}$ is the density of the neutral hydrogen, $H(z)$ is the Hubble
constant, $x_{HI}$ is the neutral fraction of hydrogen, $\delta$ is the local
baryon overdensity, $\nu_0$ is the rest frequency of the 21cm transition, and 
the remainder are the usual constants.
Plugging in the cosmological parameters from \citet{planck_et_al2015}, 
we get 
\begin{equation}\label{eqn:final_tb}
    \delta{T_{b}} \approx T_{0}\,x_{\text{\tiny{HI}}}\,(1+\delta)\, \xi ,
\end{equation}
where $\xi\equiv1-\Tcmb/\Tspin$ and $T_0\equiv26.7 \,
\text{mK} \sqrt{(1+z)/10}$.

If the spin temperature is larger than $\Tcmb$, we get the $21\,\textrm{cm}$ signal in
emission with respect to the CMB, and $\xi\sim1$. However, if $\Tspin$ is less than $\Tcmb$,
$\delta T_b$ is negative and $\xi$ can potentially become large.

As in P14,
we consider a ``weak heating" scenario in which $\Tspin$ is coupled to the gas temperature via
the Wouthuysen-Field effect \citep{wouthuysen1952,field1958,hirata2006},
but little heating has taken place prior to reionization, so that $\Tspin<\Tcmb$.
In this scenario, 
because we have assumed little heating, we can approximate $\xi$ as having negligible spatial
dependence, and therefore $T_0^2\xi^2$ becomes a simple multiplicative scalar to the 
21cm power spectrum:
\begin{equation}\label{eqn:d2pspec_model}
    \Delta^2_{21}(k) = T_0^2\xi^2(z)\Delta_{i}^{2}(k),
\end{equation}
where $\Delta_{i}^{2}(k)$ is the dimensionless HI power spectrum. 

As shown in P14, the maximum value of the
prefactor in Equation \eqref{eqn:d2pspec_model} is given by 
a no-heating scenario where the spin temperature follows the kinetic gas temperature,
which is held in equilibrium with the CMB via Compton scattering until $z_{\rm dec}\approx150$
\citep{furlanetto_et_al2006} and then cools adiabatically
as $(1+z)^2$.
In this case, $\xi$ is given by
\begin{equation}\label{eqn:maxxi}
\xi = 1 -\frac{1+z_{\rm dec}}{1+z} \approx -\frac{150}{1+z}.
\end{equation}
At $z=8.4$, this corresponds to a minimum bound on the spin temperature of $\Tspin>1.5~{\rm K}$.

We can now flip this argument around and, for a measured upper bound on $\Delta^2_{21}(k)$, we
can use models for $\Delta_i^2(k)$ in Equation \eqref{eqn:d2pspec_model} to place a bound
on $\Tspin$.  
%
We consider a class of ``patchy" reionization models (P12a;P14) which
approximates the ionization power spectrum as flat between minimum and maximum
bubble sizes, $\kmin$ and $\kmax$, respectively:
\begin{equation}\label{eqn:patchy_bound}
    \Delta^{2}_{i}(k) = (x_{\text{HI}} -
x_{\text{HI}}^{2})/\ln{(\kmax/\kmin)}.
\end{equation}
For combinations of $\kmin$ and $\kmax$,
we 
determine the minimum spin temperature 
implied by the $2\sigma$ $21\,\textrm{cm}$ power
spectrum upper limits shown in Figure \ref{fig:final_pspec}.
Figure \ref{fig:patchy_bound} shows the results of these bounds
for neutral fractions of $x_{\text{HI}}=$ 0.1, 0.3, 0.5, 0.7, and 0.9.
In almost all cases (excepting $x_{\text{HI}}=0.1,0.9$ for $\kmin<0.1\hMpci$), 
we find that $\Tspin\gtrsim 3\,\text{K}$, indicating 
that our measurements are inconsistent with the spin temperature
being coupled to a kinetic temperature governed strictly by
adiabatic expansion.

Our results become more interesting in the range of $\kmin\sim0.1$ and
$\kmax\sim30$ representative of fiducial simulations
\citep{zahn_et_al2007,lidz_et_al2008}.  For neutral fractions of 0.3, 0.5, and
0.7, we find that $\Tspin\gtrsim 4\,\text{K}$. \citet{pober_et_al2015} improves
on these results by using a simulation-based framework, rather than relying on
coarse parametrizations of the power spectrum shape.
They compare the limits they find 
to the amount of heating possible given the currently observed star
formation rates in high-redshift galaxy populations
\citep{bouwens_et_al2014,mcleod_et_al2014} and assumptions about the
relationship between star formation rates and X-ray luminosities
\citep{furlanetto_et_al2006,pritchard_loeb2008,fialkov_et_al2014}.
Assuming the
midpoint of reionization lies close to $z=8.4$ (a reasonable assumption given
that \citealt{planck_et_al2015} suggests a midpoint of $z=8.8$), both the bounds
found in this paper and \citet{pober_et_al2015} show evidence for 
heating that places constraints on the possible values for the star formation
rate/X-ray luminosity correlation given certain models of the star formation
rate density redshift evolution. We refer the reader
to \citet{pober_et_al2015} for a detailed examination of these results.

\section{Discussion}\label{sec:discussion}
%
%
%

The improvement in our results over those in P14 are the result of four
major advances:
\begin{enumerate}
\item the expansion of PAPER to 64 antennas doubled our instrument's power spectrum sensitivity,
\item using OMNICAL for redundant calibration significantly improved the clustering of measurements
over the previous implementation of LOGCAL used in P14,
\item fringe-rate filtering further improved power spectrum sensitivity by $\sim$50\% and suppressed
systematics associated with foregrounds low in the primary beam, and
\item moving from a lossless quadratic estimator targeting difference modes
in redundant measurements to an OQE (with carefully calibrated signal
loss) significantly reduced contamination from residual foregrounds.
\end{enumerate}
Figure \ref{fig:pspec_comp} illustrates the effect of some of these advances on the final
power spectrum.
Other important advances include the use of the median statistic to reduce the impact
of non-Gaussian outliers in power-spectral measurements, and the use of a Cholesky
decomposition of the Fisher information matrix to help reduce leakage 
from highly contaminated modes within the wedge.

These new techniques and improvements to calibration have reduced the measured
bias in nearly all wavebands by an order of magnitude or more. 
The use of
OMNICAL to accurately calibrate the relative complex gains of the antennas has
shown to be a major improvement to the data-reduction pipeline. The accuracy and improvement of
this calibration brings redundant baselines into impressive agreement with one another
(see Figures \ref{fig:omniview} and \ref{fig:density}),
and provides important diagnostic information for
monitoring the health of the
array, flagging RFI events, and otherwise assessing data quality.
Fringe-rate filtering, which is described in greater depth in \citep{parsons_et_al2015}, is also
proving to be a flexible and powerful tool for controlling direction-dependent gains and
improving sensitivity. 

As sensitivity improves, it will be possible to determine more accurately than
\citet{moore_et_al2015} what the actual level of polarized emission, and thus
leakage, may be.  Independent fringe-rate filtering of the XX and YY
polarizations prior to summation has the potential to better match these
polarization beams and further suppress the leakage signal if the polarized signal
turns out to be significant.

The end result is a major step forward, both for PAPER and for the field of 21cm cosmology.
While we have not yet made a detection of the 21cm cosmological signal, our limits are
now within the range of some of the brighter models.  As discussed in \citet{pober_et_al2015},
another order-of-magnitude improvement in sensitivity will make 21cm measurements highly constraining.

\section{Conclusions}\label{sec:conclusion}

We present new upper limits on the $21\,\textrm{cm}$ reionization power spectrum at $z=8.4$,
showing a factor of $\sim$4 improvement over the previous best result (P14).
We find a $2\sigma$ upper limit of \mKlimit by fitting a
flat power spectrum in a $k$ range from $0.15<k<0.5\,\hMpci$ to the
dimensionless power spectrum, $\Delta^{2}(k)$, measured by the PAPER instrument. 
We coarsely show that these upper limits imply a minimum spin
temperature for hydrogen in the IGM.  Although these limits are dependent on
the model chosen for the power spectrum, we use a patchy reionization model
to show that limits of $T_s>4\,\textrm{K}$ are fairly generic for models with
ionization fractions between 0.3 and 0.7.
A more detailed analysis of the implied constraints on spin temperature using semi-analytic reionization/heating simulations is presented in a forthcoming paper \citep{pober_et_al2015}.

The power spectrum results that we present continue to be based on
the delay-spectrum approach to foreground avoidance presented in 
P12b and first applied in P14.  The application of a delay filter over
a wide bandwidth continues to be one of the most powerful techniques yet
demonstrated for managing bright smooth-spectrum foregrounds.  In this
paper, we extend the analysis in P14 with improved fringe-rate filtering,
improved redundant calibration with OMNICAL, and with an OQE
that, while not perfectly lossless, is more adept at down-weighting residual foregrounds.
The combined effect of these improvements leaves a power-spectral measurement that
is not consistent with zero at the 4.7$\sigma$-level, which we expect is a result of
contamination from crosstalk and foregrounds.
With the expansion of PAPER to 64 antennas, the extended 135 day
observing campaign,
and the added sensitivity benefits of fringe-rate filtering, combined with
the optimization of antenna positions in PAPER for highly redundant
measurements, this thermal
noise limit is beginning to enter the realm of constraining realistic models of reionization.

Forthcoming from PAPER will be two seasons of observation with a 128-element array.
Following the same analysis as presented here, that data set is expected to improve 
over the PAPER-64 sensitivity by a factor of $\sim$4 (in mK$^2$), with the potential for another boost to sensitivity
should the new 16-m baselines provided in the PAPER-128 array configuration prove to be
usable.  There also remains the potential for further improvements to sensitivity through the
use of longer baselines, if foregrounds can be managed effectively.
As has been done recently for PAPER-32 \citep{jacobs_et_al2014,moore_et_al2015}, 
future work will also extend PAPER-64 analysis
to a range of redshifts and examine the power spectrum of polarized emission.


With recent breakthroughs in foreground management, the sensitivity 
limitations of current experiments are becoming clear.  Although collecting area is vital,
as discussed in \citet{pober_et_al2014}, the impact of collecting area
depends critically on the interplay of array configuration with foregrounds.
Despite a large spread in collecting areas between PAPER, the MWA, and LOFAR,
in the limit that foreground avoidance is the only viable strategy, these
arrays all deliver, at best, comparable low-significance detections of fiducial models
of reionization.  To move beyond simple detection, next-generation instruments must
deliver much more collecting area with very compact arrays.

The Hydrogen Epoch of Reionization Array (HERA) and the low frequency Square Kilometre Array (SKA-Low) are next generation experiments that aim to make significant detections of the $21\,\textrm{cm}$ power spectrum and begin characterizing it. SKA-Low has secured pre-construction funding for a facility in western Australia. HERA was recently granted funding for its first phase under the National Science Foundation's {\it Mid-Scale Innovations Program}. 
HERA uses a close packing of 14-m diameter
dishes designed to minimize the width of the delay-space kernel
$\tilde{A}_\tau$ in Equation \eqref{eqn:delay_transform}.
Sensitivity forecasts for a 331-element HERA array and SKA-Low
show that they can deliver detections of the 21cm reionization signal
at a significance of 39$\sigma$ and 21$\sigma$, respectively, using the same the conservative 
foreground avoidance strategy employed in this paper
\citep{pober_et_al2014}.  HERA is the natural successor
to PAPER, combining a proven experimental strategy with the 
sensitivity to deliver results that will be truly transformative for
understanding of our cosmic dawn.


\section{Acknowledgements} 

PAPER is supported by grants from the National Science Foundation (NSF; awards 0804508,
1129258, and 1125558).  A.R.P, J.C.P, and D.C.J would like to acknowledge NSF support
(awards 1352519, 1302774, and 1401708, respectively).
J.E.A would like to acknowledge a generous grant from the Mount Cuba Astronomical Association for
computing resources.
We graciously thank SKA-SA for site infrastructure and observing support. 
We also thank interns Monde Manzini and
Ruvano Casper from Durban University of Technology, who helped expand
the array from 32 to 64 antennas.
Thanks also to Josh Dillon for helpful discussions on optimal quadratic
estimators. 

\bibliographystyle{apj}
\bibliography{biblio}

\appendix
\section{ERRATUM}
\label{app:erratum}

In this erratum, we retract the upper limits on the $21\,\textrm{cm}$ power spectrum
presented in the original manuscript.  The original manuscript reported an upper
limit on $\Delta_{21}^2(k)$ of \mKlimit at $z=8.4$ in the range
$0.15<k<0.5\hMpci$.  This analysis under-estimated the level of signal loss, or attenuation of
the target cosmological $21\,\textrm{cm}$ signal associated with the chosen power spectrum
estimator, and also under-estimated the statistical error on those estimates.
A revised result, with a new analysis, is presented in \kolopaniscitet.  Below,
we briefly summarize the errors in the original analysis and how they are
corrected. For an in-depth analysis and discussion of the errors, we refer the reader to
\chengcitet.

Signal loss was expected in the original analysis because the
covariance matrices, $\textbf{C}$, used to weight the un-normalized bandpower
estimates, ${\widehat{\textbf{q}}}_\alpha$, in

\begin{equation}
{\widehat{\textbf{q}}}_{\alpha} = {\mathbf x}\textbf{C}^{-1}\textbf{Q}_\alpha \textbf{C}^{-1}{\mathbf x}
\end{equation} 

\noindent were empirically estimated from a time-averaged finite ensemble of the data,
$\mathbf x$, such that $\mathbf{C}\rightarrow \widehat{\textbf{C}}_{x}=\langle {\mathbf x} {\mathbf x}^\dagger\rangle_{t}$.
While the true covariance $\textbf{C}$ leads to an inherently unbiased lossless estimator of the power spectrum, using an empirically-estimated $\widehat{\textbf{C}}$ can lead to signal loss. Specifically, weighting data by an empirically estimated covariance carries the risk of over-fitting and down-weighting EoR fluctuations that are coupled to the data. In $\chengcitet$, it is shown that these couplings are especially strong in the fringe-rate filtered PAPER-64 dataset.

The first and most impactful error relates to the method by which signal loss
was estimated.  To assess signal loss from the empirically estimated covariance matrix, different realizations of mock
cosmological signals $\mathbf e$ of known amplitudes are added to the original data to form a new data vector, 
${\mathbf r}\equiv{\mathbf x} + {\mathbf e}$.
New covariance matrices, 
$\widehat{\textbf{C}}_r=\langle{\mathbf r}\mathbf{r}^\dagger\rangle_{t}$, 
are used to estimate un-normalized bandpowers, 
${\widehat{\textbf{q}}}_{\alpha,r}$, which can be written as 

\begin{align}
\widehat{\textbf{q}}_{\alpha,r} ={\mathbf x}\widehat{\textbf{C}}_r^{-1}\textbf{Q}_\alpha \widehat{\textbf{C}}_r^{-1}{\mathbf x}+
{\mathbf x}\widehat{\textbf{C}}_r^{-1}\textbf{Q}_\alpha \widehat{\textbf{C}}_r^{-1}{\mathbf e}+
{\mathbf e}\widehat{\textbf{C}}_r^{-1}\textbf{Q}_\alpha \widehat{\textbf{C}}_r^{-1}{\mathbf x}+
{\mathbf e}\widehat{\textbf{C}}_r^{-1}\textbf{Q}_\alpha \widehat{\textbf{C}}_r^{-1}{\mathbf e}.
\label{eq:crossterms}
\end{align}

\noindent The normalized power estimate can then be compared to the known injected power in $\mathbf{e}$ to estimate signal loss.

The key error in the previous analysis was to assume that, since $\mathbf e$ was statistically independent of $\mathbf x$, that
the two middle cross-terms in Eq. (\ref{eq:crossterms}) would
average to zero in an ensemble. 
However, as shown in $\chengcitet$ and \citet{switzer_et_al2015}, these cross-terms can contain
significant negative power because $\widehat{\textbf{C}}_r$ contains information
that correlates the two vectors. Ignoring these cross-terms leads to a significant under-estimate of signal loss.

As a result, we presented negligible signal loss in our original analysis, when in fact approximately
$\sim{99.99}\%$ of the signal was removed $\chengcitep$. Correcting for the actual signal loss is the biggest factor revising the upper limit
on $\Delta^2_{21}$.

The second mistake made in the original analysis was to under-estimate the statistical errors in the reported power spectrum
estimates.  The original analysis used a bootstrap resampling technique on power spectral measurements over the baseline and time axes.
However, fringe-rate filtering introduces significant correlations in the data
along the time axis.  As is discussed in \chengcitet, bootstrapping across correlated samples can result in a significant under-estimate
of the variation in the data if the number of resamplings is not equal to the number of independent samples in the data, as in the case 
of the original analysis. The error bars associated with this oversampling were under-estimated by approximately a factor of 2 (in $\textrm{mK}$).
The revised analysis in $\kolopaniscitet$ only applies bootstrap resampling across the baseline axis to avoid this problem.

The mistake in estimating the statistical errors should have become apparent when comparing results to our theoretical
thermal noise sensitivity.  Unfortunately, a third miscalculation was made in estimating the thermal noise sensitivity.
As detailed in \chengcitet, this miscalculation stemmed from numerous small mismatches between the idealized analysis
pipeline used to estimate sensitivity and the actual analysis applied to the data.  As a result, our estimated thermal
noise sensitivity was approximately a factor of 3 low (in $\textrm{mK}$), leading to the mistaken impression that our
errorbars were consistent with the level of thermal noise.

In summary, we retract the power spectrum results shown in Figures 18 and 20 in the original manuscript.
Results that relied on the original limits, including those presented in Figure 21,
are retracted.  Additionally, the companion paper to the original manuscript,
Pober et al. 2015, used the original limits to place constraints on the spin
temperature of the intergalactic medium (IGM) at z = 8.4.  Our revised limits
do not place significant constraints on the IGM temperature and the results of
Figure 4 from Pober et al. 2015 should be disregarded.  However, we note that
their analysis would still be relevant should a future experiment place
constraints on the 21 cm signal similar to those claimed in our original
manuscript.
An updated analysis of this same dataset is presented in $\kolopaniscitet$, where these revised results are put into context with measurements at other 
redshifts.

\end{document}